\documentclass[particles,article,accept,moreauthors]{Definitions/mdpi} 

\firstpage{1} 
\pubvolume{1}
\issuenum{1}
\articlenumber{0}
\pubyear{2024}
\copyrightyear{2024}
\datereceived{ } 
\daterevised{ } 
\dateaccepted{ } 
\datepublished{ } 
\hreflink{https://doi.org/} 

\def \be  {\begin{equation}}
\def \ee  {\end{equation}}
\def \bea {\begin{eqnarray}}
\def \eea {\end{eqnarray}}
\def \nn  {\nonumber}

\usepackage{amsmath}
\DeclareMathOperator{\Tr}{Tr}

\usepackage{mathrsfs}  

\Title{Mass Spectrum of Non-Charmed and Charmed Meson States in Extended Linear-Sigma Model}

\TitleCitation{Mass Spectrum of Non-Charmed and Charmed Meson States in Extended Linear-Sigma Model}

\Author{Azar I. Ahmadov$^{1,2,\dagger}$ and Azzah A. Alshehri $^{3,\ddagger}$ and Abdel Nasser Tawfik $^{4,\ast}$\orcidA{}}

\AuthorNames{A. Ahmadov, A. Alshehri, and A. Tawfik}

\address{%
$^{1}$ \quad  Department of Theoretical Physics, Baku State University, Z. Khalilov st. 23, AZ1148, Baku, Azerbaijan \\
$^{2}$ \quad Institute for Physical Problems, Baku State University, Z. Khalilov st. 23, AZ1148, Baku, Azerbaijan \\
$^{3}$ \quad University of Hafr Al Batin, Al Jamiah street, Hafr Al Batin 39524, KSA \\
$^{4}$ \quad Future University in Egypt (FUE), Fifth Settlement, 11835 New Cairo, Egypt} 

\corres{Correspondence: a.tawfik@fue.edu.eg}

\abstract{
The mass spectrum of different meson particles is generated using an effective Lagrangian of the extended linear-sigma model (eLSM) for scalar and pseudoscalar meson fields and quark flavors, up, down, strange, and charm. Analytical formulas for the masses of scalar, pseudoscalar, vector, and axialvector meson states are derived assuming global chiral symmetry. The various eLSM parameters are analytically deduced and numerically computed. This enables accurate estimations of the masses of sixteen non-charmed and thirteen charmed meson states at vanishing temperature. The comparison of these results to recent compilation of the particle data group (PDG) allows to draw the conclusion that the masses of sixteen non-charmed and thirteen charmed meson states calculated in the eLSM are in good agreement with the PDG. This shows that the eLSM, with its configurations and parameters, is an effective theoretical framework for determining the mass spectra of various non-charmed and charmed meson states, particularly at vanishing temperature.
}

\begin{document}

\setcounter{section}{-1} 

\section{Introduction}
\label{sec:intr}

Prior to the discovery of Quantum Chromodynamics (QCD), the theory of strong interactions, Gell-Mann and Levy proposed the linear-sigma model (LSM) as a means to introduce a field represented by a point particle. This field is confined within a fixed manifold and describes the interactions of pions \cite{Gell-Mann:1960mvl}. The field itself corresponds to a spinless sigma, the scalar meson. Several studies have been performed utilizing LSM like $\mathcal{O}(4)$ LSM \cite{Gell-Mann:1960mvl}. The utilization of LSM, which incorporates quark degrees of freedom, is made possible by the extension that incorporates the dynamical realization of the pseudoscalar and scalar mesons as a linear representation of  chiral symmetry which is weakly broken by current quark masses. As a result, the model can be utilized as an effective QCD-like model \cite{Lenaghan:1999si,Petropoulos:1998gt,PhysRevD.9.1825,PhysRevD.12.2781,PhysRevD.21.278,particles3010002}. 

Although QCD can be solved perturbatively at very high energies, only approximate numerical solutions are feasible in the nonperturbative regime \cite{Tawfik:2004sw}. Despite the significant computational expenses demanded, lattice QCD simulations do not appear to be reliable especially at finite densities. Consequently, the utilization of effective QCD-like models like eLSM becomes necessary due to their shared global symmetries with QCD and low computational efforts. The chiral symmetry is widely regarded as the first approximation in comprehending the composition of hadrons. Four decades ago, the spin-zero mass spectrum and leptonic decay constants were calculated in the one-loop approximation of the SU(4) linear-sigma model \cite{Geddes:1979nd}. eLSM explored the phenomenology of charmed mesons \cite{Eshraim2015}. Recently, the quark-hadron phase structure was explored at limited temperatures and densities using the mean-field approximation of the SU(4) Polyakov linear-sigma model (PLSM) \cite{Diab:2016iig}. The assumption is that the $N_f=4$ Lagrangian is comparable with $N_f=3$ Lagrangian. An extensive examination employing SU(3) PLSM has been undertaken to analyze the thermodynamics, phase structure, and meson masses of QCD in thermal and dense medium \cite{Tawfik:2014uka,Tawfik:2016edq}, and finite magnetic fields \cite{Tawfik:2016lih,Tawfik:2016gye,Tawfik:2019rdd,Tawfik:2021eeb}. The masses of various meson states have been calculated \cite{Tawfik:2023egf,Tawfik:2015tga,Tawfik:2014gga}. Bot only baryon density but also finite isospin asymmetry are also analyzed \cite{Tawfik:2019tkp}. As systematic comparison between the mean-field approximation and optimized perturbation theory has been reported as well \cite{Tawfik:2019kaz}. These studies encompassed a wide range of conditions and yielded valuable insights into these fundamental aspects of QCD.

In this regard, we recall that the Lagrangian of the gauge theory for color interactions of quarks and gluons, QCD, is invariant under local color transformations so that the physical content remains invariant if the colors of quarks and gluons are transformed. The interactions themselves are flavor-blind, on the other hand. In eLSM, the global chiral symmetry is explicitly violated by non-vanishing quark masses and quantum effects \cite{Giacosa:2012cq}, and spontaneously disrupted by the non-vanishing expectation value of the quark condensate in the QCD vacuum. As previously mentioned, the eLSM framework is considered as an effective approach in examining different QCD symmetries, including i) the isospin symmetry, which is a global transformation referring to SU(2) rotation in flavor space of up and down quarks, where the Lagrangian is invariant for identical or vanishing masses \cite{Tawfik:2019tkp,Tawfik:2023egf,Tawfik:2023all}, and ii) the global chiral symmetry, which is exact in the chiral limit of massless QCD, i.e., left- and right-quarks become degenerate. The chiral symmetry is broken by QCD vacuum properties, such as the Higgs mechanism, where pions are the lightest Goldstone bosons of the broken symmetry \cite{Tawfik:2015uda,Tawfik:2015tga}; iv) discrete $C$, $P$, and $T$ symmetries \cite{Eshraim2015}; and v) classical dilatation (scale) symmetry \cite{Eshraim2020}.

The present study utilizes an extended linear-sigma model (eLSM) in an effective Lagrangian in which four quark flavors along with scalar and pseudoscalar meson fields are included \cite{Diab:2016iig,Klempt:2007cp,AMSLER200461}. To ensure accurate results the integration of Polyakov-loop potential is essential, which can be derived from pure Yang-Mills lattice simulations \cite{POLYAKOV1978477,PhysRevD.20.2610,SVETITSKY1982423}. We assume that such a configuration allows for the proper estimation of various meson states $\langle \bar{q} q\rangle = \langle \bar{q}_{\ell} q_{r} - \bar{q}_{\ell} q_{r}\rangle \neq 0$ \cite{VAFA1984173}. These meson states, characterized by their chiral structures, can be categorized based on specific quantum numbers such as orbital angular momenta $J$, parity $P$, and charge conjugates $C$. This classification results in scalar mesons with $J^{PC}=0^{++}$, pseudoscalar mesons with $J^{PC}=0^{-+}$, vector mesons with $J^{PC}=1^{--}$, and axialvector mesons with $J^{PC}=1^{++}$ \cite{Tawfik:2019rdd,Tawfik:2014gga}. The present study targets a systematic analysis of various non-charmed and charmed meson states. The effective Lagrangian is utilized to describe the properties of low-lying meson states that are non-charmed and charmed, determining their mass spectra at vanishing temperature. 

At a vanishing temperature, the masses of various meson states are well measured \cite{ParticleDataGroup:2022pth}. Their excellent reproduction by means of eLSM, allows to determine its parameters so that the meson masses in thermal and dense medium can then be calculated.

The colour gauge field theory, in which a linear binding potential with one-gluon exchange corrections and four quark flavors are coupled, was utilized to determine the meson masses \cite{BARBIERI1976125}. A considerable agreement with the experimental data is concluded especially for mesons of masses larger than 1 GeV \cite{BARBIERI1976125,Parganlija:2010ac,PhysRevD.80.113001}. In two-flavour linear sigma model the meson vacuum phenomenology was studied and concluded that the inclusion of additional scalar degrees of freedom is necessary \cite{Parganlija:2010fz}. Within the $U(3)\times U(3)$ linear sigma model, the meson properties are studied and concluded that the $U_A(1)$-breaking term plays an important role in the generation of meson masses \cite{Napsuciale:1998ip}.  

The script is arranged as follows. Section \ref{sec:frml} describes the formalism of the SU(4) extended linear-sigma model's mesonic component of the Lagrangian. Section \ref{sec:Rslts} presents our findings on the mass spectra of non-charmed and charmed meson states. The last section, \ref{sec:cncl}, discusses the overall perspective.

\section{Mesonic Lagrangian of extended Linear-Sigma Model}
\label{sec:frml}

The extended linear-sigma model (eLSM) incorporates chiral symmetry in a linear representation \cite{Gell-Mann:1960mvl}. The nonlinear representation only considers Goldstone bosons but not vector mesons \cite{GASSER1984142,Bando:1987br}. The linear representation also allows us to examine scalar Goldstone bosons, but its expansion enables the introduction of vector and axialvector mesons. As mentioned in the introduction, eLSM considers chiral symmetry among other QCD symmetries \cite{Rosenzweig:1981cu}.

We assume that the Lagrangian for $N_f=4$ with global chiral invariance \cite{Diab:2016iig} is similar to the comparable Lagrangian for $N_f=3$ \cite{Parganlija:2012fy}. Only for $N_f=4$, the mass term is $- 2 \Tr[\epsilon \Phi^{\dagger} \Phi].$ must be included \cite{Giacosa:2012cq}. The motivation of the mass term can be realized from the equivalence of $SU(4) \times SU(4)$ and $SU(3) \times SU(3)$ symmetry breaking Hamiltonian \cite{MOTT1975260,PhysRevD.77.125030}. The Lagrangian of the mesonic sector, comprising scalar, pseudoscalar, vector, and axialvector meson states, as well as scalar glueball, interactions and anomalies, is constructed as follows \cite{Tawfik:2019rdd,Tawfik:2014gga}:
\bea
\mathcal{L}_{\mathtt{m}} &=& \mathcal{L}_{\mathtt{ps}} + \mathcal{L}_{\mathtt{av}}
+ \mathcal{L}_{\mathtt{int}} + \mathcal{L}_{\mathtt{anomaly}} + \mathcal{L}_{\mathtt{dilaton}} + \mathcal{L}_{\mathtt{emass}}, \label{eq:Lagrng1}
\eea
where 
\bea
\mathcal{L}_{\mathtt{ps}} &=&  \Tr\left[\left(\mathcal{D}^{\mu}\Phi\right)^{\dagger}
 \left(\mathcal{D}^{\mu}\Phi\right) - m_0^2\left(\frac{G}{G_0}\right)^2 \Tr\left(\Phi^{\dagger}\Phi\right) \right] - \lambda_1\left[\Tr\left(\Phi^{\dagger}\Phi\right)\right]^2 \nn \\
 &-&  \lambda_2\Tr\left[\left(\Phi^{\dagger}\Phi\right)\right]^2 
 + \Tr\left[H\left(\Phi+\Phi^{\dagger}\right)\right],  \\
\mathcal{L}_{\mathtt{av}} &=& - \frac{1}{4} \Tr\left[\left(L^{\mu \nu}\right)^2 + \left(R^{\mu \nu}\right)^2\right] + \Tr\left\{\left[\left(\frac{G}{G_0}\right)^2\frac{m^2}{2} + \Delta\right] \left[\left(L^{\mu}\right)^2 + \left(R^{\mu}\right)^2\right]\right\} \nn \\ 
&& - 2 i g_2 \left\{ \Tr\left(L_{\mu \nu}\left[L^{\mu},L^{\nu}\right]\right) + \Tr\left(R_{\mu \nu} \left[R^{\mu},R^{\nu}\right]\right) \right\}, \\
\mathcal{L}_{\mathtt{int}} &=&  \frac{h_1}{2} \Tr\left(\Phi^{\dagger}\Phi\right) \Tr\left[\left(L^{\mu}\right)^2 + \left(R^{\mu}\right)^2\right] + h_2 \Tr\left[\left(R^{\mu}\right)^2 \Phi + \left(L^{\mu} \Phi\right)^2\right] \nn \\
&+& 2 h_3 \Tr\left(L^{\mu} \Phi R_{\mu}\Phi^{\dagger} \right),  \\
\mathcal{L}_{\mathtt{anomaly}} &=& C\left(\det \Phi + \det\Phi^{\dagger}\right)^2 + i C_{\tilde{G}} \tilde{G} \left(\det \Phi + \det\Phi^{\dagger}\right), \\
\mathcal{L}_{\mathtt{dilaton}} &=& \frac{1}{2} \left(\mathcal{D}_{\mu} G\right)^2 - \frac{1}{4} \frac{m^2_G}{\Lambda_G^2} \left(G^4 \ln\frac{G}{\Lambda_G} - \frac{G^4}{4}\right), \\
\mathcal{L}_{\mathtt{emass}} &=& - 2 \Tr\left[\epsilon\left(\Phi^{\dagger}\Phi\right)\right],
\eea 
where $G$ is the scalar glueball, while $\tilde{G}$ is the pseudoscalar glueball. $C$ and $C_{\tilde{G}}$ are constants introduced to improve the fit of mesons and glueballs, respectively. The lowest-lying glueball mass, $m_G$, is found in the quenched approximation, with no quarks \cite{Tawfik:2014eba}. $\Lambda$ represents the QCD scaling parameter. The dilaton Lagrangian is thought to replicate the QCD trace anomaly. The field $\Phi$ is a complex $N_f \times N_f$ matrix for scalar $\sigma_a$ with $J^{PC}=0^{++}$, pseudoscalar $\pi_a$ with $J^{PC}=0^{-+}$, vector with $J^{PC}=0^{--}$, and axialvector mesons with $J^{PC}=0^{++}$ (Appendix \ref{sec:matrixphi})
\bea
\Phi &=& \sum_{a=0}^{N_f^2-1} T_a \left(\sigma_a + i \pi_a\right),
\eea
where the scalar mesons are given as {\footnotesize
\bea
T_a \sigma_a &=& \frac{1}{\sqrt{2}} \nn \\
&& \left(
\begin{tabular}{c c c c}
$\frac{\sigma_0}{2}+\frac{\sigma_3}{\sqrt{2}}+\frac{\sigma_8}{\sqrt{6}}+\frac{\sigma_{15}}{2\sqrt{3}}$ & $\frac{\sigma_1-i\sigma_2}{\sqrt{2}}$ & $\frac{\sigma_4-i\sigma_{5}}{\sqrt{2}}$ & $\frac{\sigma_9-i\sigma_{10}}{\sqrt{\sqrt{2}}}$ \\
$\frac{\sigma_1+i\sigma_2}{\sqrt{2}}$ & $\frac{\sigma_0}{\sqrt{2}}-\frac{\sigma_3}{\sqrt{2}}+\frac{\sigma_8}{\sqrt{6}}+\frac{\sigma_{15}}{2\sqrt{3}}$ & $\frac{\sigma_6-i\sigma_7}{\sqrt{2}}$ & $\frac{\sigma_{11}-i\sigma_{12}}{\sqrt{2}}$ \\
$\frac{\sigma_4+i\sigma_5}{\sqrt{2}}$ & $\frac{\sigma_6+i\sigma_7}{\sqrt{2}}$ & $\frac{\sigma_0}{2}-\sqrt{\frac{2}{3}}\sigma_8 + \frac{\sigma_{15}}{2\sqrt{3}}$ & $\frac{\sigma_{13}-i\sigma_{14}}{\sqrt{2}}$ \\
$\frac{\sigma_9+i\sigma_{10}}{\sqrt{2}}$ & $\frac{\sigma_{11}+i\sigma_{12}}{\sqrt{2}}$ & $\frac{\sigma_{13}+i\sigma_{14}}{\sqrt{2}}$ & $\frac{\sigma_0}{2}-\frac{\sqrt{3}}{2}\sigma_{15}$
\end{tabular}
\right). \hspace*{10mm} \label{eq:Tapia}
\eea }
Similarly, the pseudo-scalar mesons become {\footnotesize
\bea
T_a \pi_a &=& \frac{1}{\sqrt{2}}  \nn \\
&& \left(
\begin{tabular}{c c c c}
$\frac{\pi_0}{2}+\frac{\pi_3}{\sqrt{2}}+\frac{\pi_8}{\sqrt{6}}+\frac{\pi_{15}}{2\sqrt{3}}$ & $\frac{\pi_1-i\pi_2}{\sqrt{2}}$ & $\frac{\pi_4-i\pi_{5}}{\sqrt{2}}$ & $\frac{\pi_9-i\pi_{10}}{\sqrt{\sqrt{2}}}$ \\
$\frac{\pi_1+i\pi_2}{\sqrt{2}}$ & $\frac{\pi_0}{\sqrt{2}}-\frac{\pi_3}{\sqrt{2}}+\frac{\pi_8}{\sqrt{6}}+\frac{\pi_{15}}{2\sqrt{3}}$ & $\frac{\pi_6-i\pi_7}{\sqrt{2}}$ & $\frac{\pi_{11}-i\pi_{12}}{\sqrt{2}}$ \\
$\frac{\pi_4+i\pi_5}{\sqrt{2}}$ & $\frac{\pi_6+i\pi_7}{\sqrt{2}}$ & $\frac{\pi_0}{2}-\sqrt{\frac{2}{3}}\pi_8 + \frac{\pi_{15}}{2\sqrt{3}}$ & $\frac{\pi_{13}-i\pi_{14}}{\sqrt{2}}$ \\
$\frac{\pi_9+i\pi_{10}}{\sqrt{2}}$ & $\frac{\pi_{11}+i\pi_{12}}{\sqrt{2}}$ & $\frac{\pi_{13}+i\pi_{14}}{\sqrt{2}}$ & $\frac{\pi_0}{2}-\frac{\sqrt{3}}{2}\pi_{15}$
\end{tabular}
\right). \hspace*{10mm} \label{eq:Tasigmaa}
\eea
}

Nonvanishing external field matrices $H$, $\Delta$, and $\epsilon$ clearly break the chiral symmetry
\bea
H &=& \sum_{a=0}^{N_f^2-1} h_a T_a = h_0 T_0 + h_8 T_8 + h_{15} T_{15}, \label{eq:H-rslts}\\
\Delta &=& \sum_{a=0}^{N_f^2-1} h_a \delta_a = h_0 \delta_0 + h_8 \delta_{15} + h_a \delta_{15}, \label{eq:Delta-rslts}\\
\epsilon &=& \epsilon_c= m_c^2=\frac{1}{2} \left[m_{\chi_{c0}}^2-m_0^2 - \lambda_1\left(\sigma_x^2+\sigma_y^2\right) - 3 \sigma_c^2 \left(\lambda_1+\lambda_2\right)\right]. \label{eq:epsilon-rslts}
\eea
That the matrix $H$ must be diagonal, the large standard deviations for some mesons, especially pions, as we shall identify, can be grasped from Eq. \eqref{eq:Lagrng1}.
The generators of the group U($N_f$) are $T_a=\lambda_a/2$, where $\lambda_a$ are the Gell-Mann matrices (see Appendix \ref{sec:GellMannMtrices}. In equations \eqref{eq:H-rslts}-\eqref{eq:Delta-rslts}, only three of the 15 terms resulting from the sum are specified. The reason for this is that the matrix $H$ was selected to be diagonal. Appendix \ref{sec:matrixphi} will include the whole matrix $H$.

%

In SU(4)$_{\ell}\, \times $SU(4)$_r$ model, the quark condensates are given as
\bea
\sigma_x &=& \frac{\sigma_0}{\sqrt{2}} + \frac{\sigma_8}{\sqrt{3}} + \frac{\sigma_{15}}{\sqrt{6}}, \\
\sigma_y &=& \frac{\sigma_0}{2} - \sqrt{\frac{2}{3}} \sigma_8 + \frac{1}{2\sqrt{3}} \sigma_{15}, \\
\sigma_c &=& \frac{\sigma_0}{2}  + \frac{\sqrt{3}}{2} \sigma_{15},
\eea
where $\sigma_x$ represents the condensate of light quarks (up and down), whereas $\sigma_y$ represents the strange quark condensate. $\sigma_c$ represents the charm quark condensates. The complex $N_f \times N_f$ matrix for scalar $J^{PC}=0^{++}$, pseudoscalar $J^{PC}=0^{-+}$, vector $J^{PC}=0^{--}$, and axialvector mesons $J^{PC}=0^{++}$ corresponds to $\Phi$, Eq. (\ref{eq:phimatrix}). Using $m_0=(g/2)\Phi$, where $g$ represents the Yukawa coupling, the quark masses can be related the quark condensates
\bea
m_u &=& \frac{g}{2}\left[\frac{\sigma_0}{\sqrt{2}} + \frac{\sigma_8}{\sqrt{3}}  + \frac{\sigma_{15}}{\sqrt{6}}\right]= \frac{g}{2} \sigma_x, \\
m_d &=& \frac{g}{2}\left[\frac{\sigma_0}{\sqrt{2}} + \frac{\sigma_8}{\sqrt{3}}  + \frac{\sigma_{15}}{\sqrt{6}}\right]= \frac{g}{2} \sigma_x, \\
m_s &=& \frac{g}{2}\left[\frac{\sigma_0}{\sqrt{2}} - \frac{2\sigma_8}{\sqrt{3}}  + \frac{\sigma_{15}}{\sqrt{6}}\right]=\frac{g}{\sqrt{2}} \sigma_y, \\
m_c &=& \frac{g}{2}\left[\frac{\sigma_0}{\sqrt{2}} - \sqrt{\frac{3}{2}} \sigma_{15}\right]=\frac{g}{\sqrt{2}} \sigma_c.
\eea

The covariant derivative of the scalar mesons is expressed as
\bea
\mathcal{D}^{\mu}\Phi &=& \delta_{\mu}\Phi - i g_1 \left(L^{\mu}\Phi - \Phi R^{\mu}\right) - i e A^{\mu}\left[T_3,\Phi\right],
\eea
where $A_{\mu} =g A_{\mu}^a \lambda^a/2$ is the electromagnetic field. For vector and axialvector meson nonets, we have
\bea
L^{\mu\nu} &=& \delta_{\mu}L^{\nu} - i e A^{\mu}\left[T_3, L^{\nu}\right] - \left\{\delta^{\nu}\mathcal{L}^{\mu}-i e A^{\nu}\left[T_3,L^{\mu}\right]\right\}, \\
R^{\mu\nu} &=& \delta_{\mu}R^{\nu} - i e A^{\mu}\left[T_3, R^{\nu}\right] - \left\{\delta^{\nu}R^{\mu}-i e A^{\nu}\left[T_3,R^{\mu}\right]\right\}, 
\eea
where $L^{\mu}=\sum_{a=0}^{N_f^2-1} T_a (V_a^{\mu} + A_a^{\mu})$ and $R^{\mu}=\sum_{a=0}^{N_f^2-1} T_a (V_a^{\mu} - A_a^{\mu})$.
The remaining quantities can be deduced as follows:
\bea
\Tr\left(\Phi^{\dagger}\Phi\right) &=& \frac{1}{4} \left[\left(\sigma_x\right)^2 + \left(\sigma_y\right)^2 + \left(\sigma_{c}\right)^2 \right], \\
\Tr \left[\epsilon \Phi^{\dagger} \Phi\right] &=& \frac{1}{2} \epsilon_c \left(\sigma_c\right)^2, \\
\Tr \left[H \left(\Phi^{\dagger} + \Phi\right)\right] &=& h_x \sigma_x + h_y \sigma_y + h_c \sigma_c,\\
C\left(\det \Phi + \det \Phi^{\dagger} \right) &=& \frac{C}{4}\left(\sigma_x\right)^2 \sigma_y \sigma_c, 
\eea
\bea
\left[\Tr\left(\Phi^{\dagger}\Phi\right)\right]^2 &=& \frac{1}{4} \left[
\left(\sigma_x\right)^4 + \left(\sigma_y\right)^4 + \left(\sigma_c\right)^4 + 2 \left(\sigma_x\right)^2 \left(\sigma_y\right)^2 + 2 \left(\sigma_y\right)^2 \left(\sigma_c\right)^2 + 2 \left(\sigma_x\right)^2 \left(\sigma_c\right)^2
\right]. \hspace*{7mm} 
\eea

From Eqs. (\ref{eq:Tapia}) and (\ref{eq:Tasigmaa}), we can now determine various meson states,
\bea
&& a_0^{\pm} \equiv \frac{\sigma_1\mp i\sigma_2}{\sqrt{2}},\qquad  \qquad a_0^0 \equiv \sigma_3, \nn \\
&& k^{\pm} \equiv \frac{\pi_4 \mp i \pi_5}{\sqrt{2}}, \qquad \qquad k^0 \equiv \pi_6 - i \pi_7, \qquad \qquad \bar{k^0} \equiv \frac{\pi_6 + i \pi_7}{\sqrt{2}}, \nn \\
&& \sigma_s = \frac{\sigma_0}{2} - \sqrt{\frac{2}{3}} \sigma_8 + \frac{\sigma_{15}}{2\sqrt{3}}, \qquad \qquad 
\chi_{co} = \frac{\sigma_0}{2} - \frac{\sqrt{3}}{2}\sigma_{15}, \nn \\
&& \eta_s = \frac{\pi_0}{2} - \sqrt{\frac{2}{3}}\pi_8 + \frac{\pi_{15}}{2\sqrt{3}}, \qquad \qquad 
\eta_c = \frac{\pi_0}{2} - \frac{\sqrt{3}}{2}\pi_{15}, \nn \\
&& D^0 \equiv \frac{\pi_0+i\pi_{10}}{\sqrt{2}}, \qquad  \qquad \bar{D}^0 \equiv \frac{\pi^0-i\pi_{10}}{\sqrt{2}}, \nn \\
&& D^{\pm} \equiv \frac{\pi_{11}\pm i\pi_{12}}{\sqrt{2}}, \qquad \qquad D^{\pm}_s \equiv \frac{\pi_{13}\pm i\pi_{14}}{\sqrt{2}}, \nn \\
&& D^{0}_0 \equiv \frac{\sigma_9+ i\sigma_{10}}{\sqrt{2}}, \qquad \qquad \bar{D}^{0}_0 \equiv \frac{\sigma_{9} - i\pi_{10}}{\sqrt{2}}, \nn \\
&& D^{\pm}_0 \equiv \frac{\sigma_{11}\pm i\sigma_{12}}{\sqrt{2}}, \qquad  \qquad \bar{D}^{\pm}_{s,0} \equiv \frac{\sigma_{13}\pm i\pi_{14}}{\sqrt{2}}, \nn \\
%
&& \frac{1}{\sqrt{2}}\left(\sigma_N+a_0^0\right) =\frac{a_0^0}{\sqrt{2}} + \frac{\sigma_0}{2} + \frac{\sigma_8}{\sqrt{6}} + \frac{\sigma_{15}}{2\sqrt{3}}, \nn \\
&& \frac{1}{\sqrt{2}}\left(\sigma_N-a_0^0\right) = -\frac{a_0^0}{\sqrt{2}} + \frac{\sigma_0}{2} + \frac{\sigma_8}{\sqrt{6}} + \frac{\sigma_{15}}{2\sqrt{3}}, \nn \\
&& \frac{1}{\sqrt{2}}\left(\eta_N+\pi^0\right) =\frac{\pi^0}{\sqrt{2}} + \frac{\pi_0}{2} + \frac{\pi_8}{\sqrt{6}} + \frac{\pi_{15}}{2\sqrt{3}}, \nn \\ 
&& \frac{1}{\sqrt{2}}\left(\eta_N-\pi^0\right) = - \frac{\pi^0}{\sqrt{2}} + \frac{\pi_0}{2} + \frac{\pi_8}{\sqrt{6}} + \frac{\pi_{15}}{2\sqrt{3}}. \nn
\eea
It is obvious that, for example, the values and standard deviations of $a_0^0$, as shall be listed in the next section, are determined by $\sigma_3$.

Section \ref{sec:massSpect} introduced analytical expressions for the mass spectra of non-charmed and charmed meson states.

\section{Results}
\label{sec:Rslts}
\subsection{Mass Spectra of Non-Charmed and Charmed Meson States}
\label{sec:massSpect}

At vanishing temperature, the mass spectra of non-charmed meson states can be classified into:
\begin{itemize}
\item Pseudoscalar mesons

\bea
m_{\pi}^2 &=& Z_{\pi}^2\left[m_0^2 + \left(\lambda_1+\frac{1}{2}\lambda_2\right)\sigma_x^2+\lambda_1 \sigma_y^2 + \lambda_1 \sigma_c^2
\right], \\
m_{K}^2 &=& Z_{K}^2\left[m_0^2 + \left(\lambda_1+\frac{1}{2}\lambda_2\right)\sigma_x^2-\frac{1}{2}\lambda_2 \sigma_x \sigma_y 
+ \lambda_1 \left[\sigma_y^2+\sigma_c^2\right] + \lambda_2 \sigma_y^2
\right],  \\
m_{{\eta}_{N}}^2 &=& Z_{\pi}^2\left[m_0^2 + \left(\lambda_1+\frac{1}{2}\lambda_2\right)\sigma_x^2+\lambda_1 \left[\sigma_y^2+ \sigma_c^2\right] + \frac{C}{2} \left(\sigma_x^2 + \sigma_y^2 + \sigma_c^2\right) \right], \\
%
m_{{\eta}_{S}}^2 &=& Z_{{\eta}_{S}}^2 \left[m_0^2 + \lambda_1\left(\sigma_x^2+\sigma_y^2+\sigma_c^2\right) + \lambda_2 \sigma_y^2 + \frac{C}{8} \left(\sigma_x^2 + \sigma_x^2 + \sigma_c^2\right) \right],
\eea
where, $Z_{\pi}$, $Z_{K}$, $Z_{{\eta}_{S}}$, the various wavefunction renormalization factors, are listed in Appendix \ref{sec:wavefrenorfact}.
\item Scalar mesons

\bea
m_{a_{0}}^2 &=& m_{0}^2 + \lambda_1\left[\sigma_x^2+\sigma_y^2+\sigma_c^2\right] + \frac{3}{2} \lambda_2\sigma_x^2, \\
m_{k^*_{0}}^2 &=& Z_{k^*_{0}}^2\left[m_0^2 + \left(\lambda_1+\frac{1}{2}\lambda^2\right)\sigma_x^2 
+ \frac{1}{\sqrt{2}}\lambda_2\sigma_x\sigma_y + \lambda_1\left[\sigma_y^2+\sigma_c^2\right] + \lambda_2 \sigma_c^2
\right],  \\
m_{{\sigma}_{N}}^2 &=& m_{0}^2 + 3\left(\lambda_1+\frac{1}{2} \lambda_2\right)\sigma_x^2+\lambda_1\left[\sigma_y^2+\sigma_c^2\right], \\
m_{{\sigma}_{S}}^2 &=& m_{0}^2 + \lambda_1\left[\sigma_x^2+\sigma_c^2\right] + 3 \left(\lambda_1+\lambda_2\right) \sigma_y^2,
\eea
where $Z_{k^*_{0}}$ is another wavefunction renormalization factor. Although, $(\det\Phi+ \det\Phi^{\dagger})^2$ is the anomaly term, we use in the present derivatives, the other possible anomaly terms $(\det\Phi+\det\Phi^{\dagger})$ and $(\det\Phi-\det\Phi^{\dagger})^2$ are also conjectured to affect the scalar masses, especially the earlier term.

\item Vector mesons

\bea
m^2_{\omega_N} &=& m_1^2 - m_0^2  + \frac{1}{2} \left(h_1+h_2+h_3\right) \sigma_x^2 + \frac{1}{2}h_1\left[\sigma_y^2+\sigma_c^2\right] + 2 \delta_x, \\
m^2_{\omega_S} &=& m_1^2 - m_0^2  + \frac{1}{2} h_1 \left[\sigma_x^2 + \sigma_c^2\right] + \frac{1}{2}\left(h_1+2h_2+2h_3\right) \sigma_y^2 +  2 \delta_x,  \\
m^2_{K^*} &=& m_1^2 - m_0^2 + \frac{1}{4} \sigma_x^2 
\left[g_1^2 + 2h_1 + h_2\right] +\frac{1}{\sqrt{2}} \sigma_x \sigma_y\left[h_3-g_1^2\right] \nn \\
&+& \frac{1}{2} \sigma_y^2 \left[g_1^2 + h_1 + h_2\right]+\frac{1}{2}h_1  \sigma_c^2 + \delta_x + \delta_y, \\
m^2_{\rho} &=& m^2_{\omega_N}.
\eea

\item Axialvector mesons

\bea
m^2_{a_1} &=&  m_1^2 - m_0^2  + g_1^2 \sigma_x^2 + \frac{1}{2} h_1 \left[\sigma_y^2+\sigma_c^2\right] +  \frac{1}{2} \sigma_x^2\left[h_1+h_2+h_3\right]+2\delta_x, \\
m^2_{f_{1s}} &=&  m_1^2 - m_0^2  + \frac{1}{2} h_1 \left[\sigma_x^2 + \sigma_c^2\right] + 2 g_1^2 \sigma_c^2 + \frac{1}{2}\left[h_1 + 2h_2 -2h_3\right] \sigma_y^2 + 2 \delta_y, \\
m^2_{k_{1}} &=&  m_1^2 - m_0^2  + \frac{1}{4}\sigma_x^2\left[g_1^2 + 2h_1+h_2\right] + \frac{1}{2}\sigma_y^2\left[g_1^2 + h_1 + h_2\right] \nn \\
&+& \frac{1}{\sqrt{2}} \sigma_x \sigma_y \left[g_1^2-h_3\right] + \frac{1}{2} h_1\sigma_c^2 + \delta_x + \delta_y, \\
m^2_{1N} &=& m^2_{a_1}.
\eea
\end{itemize}

Also, we determine the mass spectra of the charmed meson states.
\begin{itemize}
\item Pseudoscalar charmed mesons

\bea
m_D^2 &=& Z_D^2\left[m_0^2+\left(\lambda_1+\frac{1}{2}\lambda_2\right)\sigma_x^2 + \lambda_1\sigma_y^2+\left(\lambda_1+\lambda_2\right)\sigma_c^2 - \frac{1}{\sqrt{2}}\lambda_2 \sigma_x \sigma_y + \epsilon_c
\right], \\
m_{\eta_c}^2 &=& Z_{\eta_c}^2\left[m_0^2+\lambda_1\left[\sigma_x^2 + \sigma_y^2\right]+\left(\lambda_1+\lambda_2\right)\sigma_c^2 - \frac{C}{8}\left(\sigma_x^2 + \sigma_y^2\right) + 2\epsilon_c \right], \\
%
m_{D_s}^2 &=& Z_{D_s}^2\left[m_0^2+\lambda_1 \sigma_x^2 + \left(\lambda_1+\lambda_2\right)\sigma_y^2+\left(\lambda_1+\lambda_2\right)\sigma_c^2 - \lambda_2\sigma_y\sigma_c + \epsilon_c
\right],
\eea
where $Z_D$, $Z_{\eta_c}$ and $Z_{D_s}$ are wavefunction renormalization factors.

\item Scalar charmed mesons

\bea
m_{{\chi}_{c0}}^2 &=& m_0^2+\lambda_1 \left[\sigma_x^2+\sigma_y^2 \right] +3\left(\lambda_1+\lambda_2\right)\sigma_c^2+2\epsilon_c, \\
m_{{D}_{0}^*}^2 &=& Z_{D_0^*}^2\left[m_0^2+\left(\lambda_1+\frac{1}{2}\lambda_2\right) \sigma_x^2+\lambda_1\sigma_y^2 + \frac{1}{\sqrt{2}}\lambda_2\sigma_x\sigma_c + \left(\lambda_1+\lambda_2\right)\sigma_c^2+\epsilon_c
\right], \\
m_{{D}_{0}^{*0}}^2 &=& Z_{D_0^{*0}}^2\left[m_0^2+\left(\lambda_1+\frac{1}{2}\lambda_2\right) \sigma_x^2+\lambda_1\sigma_y^2 + \frac{1}{\sqrt{2}}\lambda_2\sigma_x\sigma_c + \left(\lambda_1+\lambda_2\right)\sigma_c^2+\epsilon_c
\right],\\
m_{{D}_{s0}^{*}}^2 &=& Z_{D_{s0}^{*}}^2\left[m_0^2+\lambda_1\sigma_x^2 + \left(\lambda_1+\lambda_2\right) \sigma_y^2 + \lambda_2\sigma_y\sigma_c+\left(\lambda_1+\lambda_2\right)\sigma_c^2+\epsilon_c
\right],
\eea
where $Z_{D_0^*}$, $Z_{D_0^{*0}}$ and $Z_{D_{s0}^{*}}$ are additional wavefunction renormalization factors.

\item Vector charmed mesons

\bea
m_{D^{*}}^2 &=& m_1^2-m_0^2+\frac{1}{4}\left(g_1^2+2h_1+h_2\right)\sigma_x^2 + \frac{1}{\sqrt{2}} \sigma_x\sigma_y\left[h_3-g_1^2\right] \nn \\
&+& \frac{1}{2}\left(g_1^2+h_1+h_2\right)\sigma_c^2+\frac{1}{2}h_1\sigma_y^2+\delta_x+\delta_c,  \\
m_{J/\psi}^2 &=& m_1^2-m_0^2+\frac{1}{2}h_1\left[\sigma_x^2+\sigma_y^2\right]+\frac{1}{2}\left(h_1+2h_2+2h_3\right)\sigma_c^2 +2\delta_c, \\
m_{D_s^*}^2 &=& m_1^2-m_0^2+\frac{1}{2}\left(g_1^2+h_1+h_2\right) \left[\sigma_y^2+\sigma_c^2\right]+\frac{1}{2}h_1\sigma_x^2  \nn \\
&+& \left(h_3-g_1^2\right)\sigma_y\sigma_c + \delta_y + \delta_c.
\eea

\item Axialvector charmed mesons

\bea
m_{D_{s1}}^2 &=& m_1^2-m_0^2+\frac{1}{2}\left(g_1^2+h_1+h_2\right) \left[\sigma_y^2+\sigma_c^2\right]+\sigma_y\sigma_c\left(g_1^2-h_3\right) + \frac{1}{2}h_1\sigma_x^2+\delta_y + \delta_c, \hspace*{7mm}\\
m_{D_1}^2 &=& m_1^2-m_0^2+\frac{1}{4}\left(g_1^2+2h_1+h_2\right) \sigma_x^2+\frac{1}{2}\left(g_1^2+h_1+h_2\right) \sigma_c^2 + \frac{1}{\sqrt{2}}\left(g_1^2-h_3\right)\sigma_x\sigma_c \nn \\
&+& \frac{1}{2}h_1 \sigma_y^2 + \delta_y + \delta_c, \\
m_{{\chi}_{c1}}^2 &=& m_1^2-m_0^2+\frac{1}{2}h_1\left[\sigma_x^2+\sigma_y^2\right]+2g_1^2 \sigma_c^2 + \frac{1}{2}\left(h_1+2h_2-2h_3\right) \sigma_c^2  + 2\delta_c.
\eea

\end{itemize}

The Lagrangian of the extended linear-sigma model, Eq. \eqref{eq:Lagrng1}, has various parameters including $h_1$, $h_2$, $h_3$, $\delta_u$, $\delta_d$, $\delta_s$, $\delta_c$, $\delta_x$, $\delta_y$, $\epsilon_u$, $\epsilon_d$, $\epsilon_s$, $\epsilon_c$, $C$, $\lambda_1$, $\lambda_2$, $g_1$, and $g_2$. All these parameters are explained and determined in the Appendix \ref{sec:Parameters}. For large $N_c$, both parameters $h_1$ and $\lambda_1$ vanish, especially for SU(3) meson masses. The other set of parameters, $\sigma_u$, $\sigma_d$, $\sigma_s$, $\sigma_c$, $m_u$, $m_d$, $m_s$, $m_c$, $f_{\pi}$, and $f_K$, can be fixed from recent compilation of particle data group \cite{ParticleDataGroup:2022pth}. 

The mass of any particle can be determined by employing the parameters presented in Tab. \ref{Tab:1} and Table \ref{Tab:1}, along with their corresponding expressions.

\begin{center}
\begin{table}[h!]
\begin{tabular}{| c | c | c | c | c | c | c | c |} \hline 
 \multirow{2}{*}{\rotatebox[origin=c]{0}{Meson}} & 
 \multicolumn{2}{c}{Mass [MeV]} & 
 \multirow{2}{*}{\rotatebox[origin=c]{0}{$h_1$}} & 
 \multirow{2}{*}{\rotatebox[origin=c]{0}{$\lambda_1$}} & 
 \multirow{2}{*}{\rotatebox[origin=c]{0}{$\lambda_2$}} & 
 \multirow{2}{*}{\rotatebox[origin=c]{0}{$g_1$}} &
 \multirow{2}{*}{\rotatebox[origin=c]{0}{$\%$ Error}}  \\ \cline{2-3} 
 & This Work & PDG & \hfill & \hfill & \hfill & \hfill & \\ \hline \hline
 \multicolumn{8}{c}{Pseudoscalar} \\ \hline
 $\pi$ & $130.58 \pm 15 $ & $139.5704\pm 0.0002$ & $5.65$ & $13.49$ & $68.30$ & $5.50$ & $0.047$ \\ 
 $K$ & $496.90\pm 4$ & $493.68 \pm 0.016$ & $5.65$ & $0.0$ & $42.0$ & $2.3$ & $0.648$ \\  
 $\eta_N$ & $1458.91\pm 21$ & $1475\pm 4$ & $5.65$ & $13.49$ & $68.30$ & $7.60$ & $1.103$ \\ 
 $\eta_S$ & $1477.94\pm 21$ & $1475\pm 4$ & $5.65$ & $13.49$ & $68.30$ & $1.25$ & $0.199$ \\ \hline \hline
 \multicolumn{8}{c}{Scalar} \\ \hline
 $a_0$ & $1482.64 \pm 15 $ & $1474\pm 19$ & $5.65$ & $4.0$ & $68.30$ & $1.0$ & $0.583$ \\ 
 $K_0^*$ & $893.40 \pm 4 $ & $891.66\pm 0.26$ & $5.65$ & $1.0$ & $25.0$ & $1.8$ & $0.195$ \\ 
 $\sigma_N$ & $1206.77 \pm 2 $ & $1200-1500$ & $5.65$ & $1.0$ & $55.0$ & $7.60$ & $0.561$ \\ 
 $\sigma_S$ & $1720.46 \pm 3 $ & $1720\pm20$ & $5.65$ & $1.0$ & $46.0$ & $7.60$ & $0.027$ \\ \hline \hline
 \multicolumn{8}{c}{Vector} \\ \hline
 $\omega_N$ & $784.42\pm 6$ & $782.65 \pm 0.12$ & $-21.5$ & $13.49$ & $68.30$ & $7.60$ & $0.225$ \\  
 $\omega_S$ & $1020.26\pm 5$ & $1019.46 \pm 0.11$ & $-16.21$ & $13.49$ & $68.30$ & $7.60$ & $0.078$ \\
 $K^*$ & $ 1390.64\pm 20$ & $ 1418 \pm 15 $ & $5.65$ & $13.49$ & $68.30$ & $7.60$ & $1.967$ \\ 
 $\rho$ & $784.42\pm 14$ & $775.5\pm 38.8$ & $-21.5$ &  $13.49$ & $68.49$ & $7.60$ & $1.137$ \\ \hline \hline
 \multicolumn{8}{c}{Axialvector} \\ \hline
 $a_1$ & $1289.17\pm21$ & $1230 \pm 30$ & $1.0$ & $1.8$ & $25$ & $1.8$ & $4.589$ \\ 
  $1f$ & $1427.26\pm2$ & $1426.4 \pm 0.9$ & $5.65$ & $1.8$ & $25$ & $2.3$ & $0.061$ \\  
  $K_1$ & $1400.61\pm 5$ & $1403 \pm 7$ & $5.65$ & $1.8$ & $25$ & $2.67$ & $0.171$\\  
  $1N$ & $1450.66\pm 5$ & $1453.4 \pm 27$ & $5.65$ & $1.8$ & $25$ & $3.5$ & $0.189$ \\ \hline 
\end{tabular}
\caption{Masses of sixteen non-charmed mesons compared with the recent compilation of the particle data group (PDG) \cite{ParticleDataGroup:2022pth}. \label{Tab:1}}
\end{table}
\end{center}

\begin{center}
\begin{table}[h!]
\begin{tabular}{| c | c | c | c | c | c | c | c |} \hline  
 \multirow{2}{*}{\rotatebox[origin=c]{0}{Meson}} & 
 \multicolumn{2}{c}{Mass [MeV]} & 
 \multirow{2}{*}{\rotatebox[origin=c]{0}{$h_1$}} & 
 \multirow{2}{*}{\rotatebox[origin=c]{0}{$\lambda_1$}} & 
 \multirow{2}{*}{\rotatebox[origin=c]{0}{$\lambda_2$}} & 
 \multirow{2}{*}{\rotatebox[origin=c]{0}{$g_1$}} &
 \multirow{2}{*}{\rotatebox[origin=c]{0}{$\%$ Error}}  \\ \cline{2-3} 
 & This Work & PDG & \hfill & \hfill & \hfill & \hfill & \\ \hline \hline
 \multicolumn{8}{c}{Pseudoscalar} \\ \hline 
 $D$ & $1868.92\pm 2 $ & $1869.6\pm 0.15$ & $5.65$ & $4.2$ & $68.30$ & $1.0$ & $0.036$ \\ 
 $\eta_s$ & $2977.46\pm 3$ & $2981 \pm 1.1$ & $5.65$ & $13.49$ & $68.30$ & $2.9$ & $0.119$ \\ 
 $D_s$ & $1965.99\pm 21$ & $1968.49\pm 0.32$ & $5.65$ & $3.6$ & $68.30$ & $1.0$ & $0.127$ \\ \hline \hline
 \multicolumn{8}{c}{Scalar} \\ \hline
 $\chi_{c0}$ & $3448.23\pm 5$ & $3414.75\pm 0.3$ & $5.65$ & $13.49$ & $68.30$ & $7.60$ & $0.971$ \\ 
 $D_0^*$ & $2421.82\pm 7$ & $2403.38\pm 14$ & $5.65$ & $1.0$ & $68.30$ & $2.0$ & $0.761$ \\ 
 $D_0^{*0}$ & $2332.02\pm 8$ & $2317.8\pm 0.06$ & $5.65$ & $1.0$ & $62.0$ & $1.0$ & $0.610$ \\ 
 $D_{s0}^*$ & $2121.6\pm 6$ & $2112.3\pm 0.3$ & $5.65$ & $1.0$ & $52.0$ & $1.0$ & $0.438$ \\ \hline \hline
 \multicolumn{8}{c}{Vector} \\ \hline
 $D^{0}$ & $2016.07\pm 3$ & $2010.28\pm0.13$ & $5.65$ &  $13.49$  & $68.30$ & $26.5$ & $0.278$ \\ 
 $J/\psi$ & $3096.15\pm 12$ & $3096.92\pm 0.011$ & $183.0$ &  $13.49$  & $68.30$ & $7.6$ & $0.025$ \\  
 $D_s^*$ & $2109.98\pm11$ & $2112.3\pm 0.5$ & $63.0$ &  $13.49$  & $68.30$ & $7.6$ & $0.120$ \\ \hline \hline
 \multicolumn{8}{c}{Axialvector} \\ \hline 
 $D_{s1}$ & $3001.98\pm 6$ & $2997.4\pm 0.17$ & $98.0$ &  $13.49$  & $68.30$ & $7.6$ & $0.152$ \\ 
 $D_{1}$ & $2846.77\pm 6$ & $2841.8\pm 0.48$ & $98.0$ &  $13.49$  & $68.30$ & $7.6$ & $0.174$ \\ 
 $\chi_{sc1}$ & $3510.57\pm 5$ & $3510.60\pm 0.7$ & $170.0$ &  $13.49$  & $68.30$ & $7.6$ & $0.001$ \\ \hline 
\end{tabular}
\caption{Masses of thirteen charmed mesons compared with the recent compilation of the particle data group (PDG) \cite{ParticleDataGroup:2022pth}. \label{Tab:2}}
\end{table}
\end{center}

Our calculations for the masses of various meson states are summarized in Tabs. \ref{Tab:1} and \ref{Tab:2}. Our calculations, in which $h_1$, $\lambda_1$, $\lambda_2$ and $g_1$ are the only free parameters, are in a good agreement with the recent compilation of the particle data group (PDG) \cite{ParticleDataGroup:2022pth}. The percent error is listed in the last column.
It measures how different the two values, namely the calculated and measured masses, are in a form of a percentage of their corresponding average value. Concretely, the percent error quantifies how close the calculated value is to that measured value. Thus, the percent error is conjectured to provide enough information for the certainty assessment. The reason why some of these free parameters play crucial roles in some meson states is to be found in the corresponding analytical expressions of such meson states.

The section that follows is devoted to the final conclusions and outlook.

\section{Conclusions and Outlook} 
\label{sec:cncl}

We have constructed the meson states $\langle \bar{q} q\rangle = \langle \bar{q}_{\ell} q_{r} - \bar{q}_{\ell} q_{r}\rangle \neq 0$ with and without charm quarks 
from the effective Lagrangian of the extended linear-sigma model. With respect to their quantum numbers, orbital angular momenta $J$, parity $P$, and charge conjugates $C$, the meson states could be classified into pseudoscalar $J^{PC}=0^{-+}$, scalar $J^{PC}=0^{++}$, vector $J^{PC}=1^{--}$, and axialvector $J^{PC}=1^{++}$. We have introduced analytical expressions for the mass spectrum of sixteen non-charmed and thirteen charmed meson states, including pseudoscalar $J^{PC}=0^{-+}$, scalar $J^{PC}=0^{++}$, vector $J^{PC}=1^{--}$, and axialvector $J^{PC}=1^{++}$. The Appendix gives further details needed for the analytical analysis. The numerical analysis and the corresponding free parameters are summarized in Tab. \ref{Tab:1} and \ref{Tab:2}, in which the calculations are confronted with the recent compilation of PDG. We conclude that the masses of the sixteen non-charmed and thirteen charmed meson states are in excellent agreement with PDG. Therefore, we also conclude that the eLSM with its set of parameters represents a suitable theoretical approach for the mass spectrum of non-charmed and charmed meson states, at vanishing temperature. The temperature dependence of these meson states represent the natural outlook to be accomplished elsewhere. Also, the in-medium modifications of these meson masses either in dense or magnetic or electric medium shall be derived elsewhere.

The heavy nonet of pseudoscalar states, which includes $\pi(1300)$, $K$, and $\eta(1475)$, is of great theoretical interest. These states seem to challenge the established principles of quark and hadron models \cite{Feng:2008zz}, which typically involve $\pi$, $K$, $\eta(547)$, and $\eta^{\prime}(958)$. Future research will involve comparing the eLSM with other approaches that consider radially excited mesons like $\pi(1300)$, $K(1460)$, $\eta(1295)$, and $\eta(1405-1475)$ \cite{Freund1968,PhysRevD.10.2960}. The applicability of U(4) to all types of interactions remains uncertain. Analysis of decays as reported in the literature \cite{Eshraim2015} indicates that U(4) may hold for dominant interactions at large-$N_c$. Further exploration using eLSM to study decay channels of various meson states is recommended.


\appendixtitles{yes} 
\appendixstart
\appendix

\section{Gell-Mann Matrices in SU(4)}
\label{sec:GellMannMtrices}

The special unitary group, SU(4), can be represented as \cite{Sbaih2013}
\bea
\mathtt{SU(4)} &=& \left\{\mathtt{A}=4\times 4\; \mathtt{complex}\; \mathtt{matrix}\; |\; \mathtt{A}^{\dagger}\mathtt{A}=1,\; \det(\mathtt{A})=1\right\}.
\eea
Analogous to the Pauli, SU(2) \cite{Arfken1985}, and Gell--Mann matrices, SU(3) \cite{Gell-Mann:1960mvl}, the hermitian matrix generators of SU(4) are defined as
\bea
\lambda_1 = \left(
\begin{tabular}{c c c c}
$0$ & $1$ & $0$ & $0$ \\
$1$ & $0$ & $0$ & $0$ \\
$0$ & $0$ & $0$ & $0$ \\
$0$ & $0$ & $0$ & $0$
\end{tabular}
\right),  
\quad && \quad
\lambda_2 = \left(
\begin{tabular}{c c c c}
$0$ & $-i$ & $0$ & $0$ \\
$i$ & $0$ & $0$ & $0$ \\
$0$ & $0$ & $0$ & $0$ \\
$0$ & $0$ & $0$ & $0$
\end{tabular}
\right),   \nn \\
\lambda_3 = \left(
\begin{tabular}{c c c c}
$1$ & $0$ & $0$ & $0$ \\
$0$ & $-1$ & $0$ & $0$ \\
$0$ & $0$ & $0$ & $0$ \\
$0$ & $0$ & $0$ & $0$
\end{tabular}
\right),  
\quad && \quad
\lambda_4 = \left(
\begin{tabular}{c c c c}
$0$ & $0$ & $1$ & $0$ \\
$0$ & $0$ & $0$ & $0$ \\
$1$ & $0$ & $0$ & $0$ \\
$0$ & $0$ & $0$ & $0$
\end{tabular}
\right),  \nn \\
\lambda_5 = \left(
\begin{tabular}{c c c c}
$0$ & $0$ & $-i$ & $0$ \\
$0$ & $0$ & $0$ & $0$ \\
$i$ & $0$ & $0$ & $0$ \\
$0$ & $0$ & $0$ & $0$
\end{tabular}
\right),  
\quad && \quad
\lambda_6 = \left(
\begin{tabular}{c c c c}
$0$ & $0$ & $0$ & $0$ \\
$0$ & $0$ & $1$ & $0$ \\
$0$ & $1$ & $0$ & $0$ \\
$0$ & $0$ & $0$ & $0$
\end{tabular}
\right),   \nn \\
\lambda_7 = \left(
\begin{tabular}{c c c c}
$0$ & $0$ & $0$ & $0$ \\
$0$ & $0$ & $-i$ & $0$ \\
$0$ & $i$ & $0$ & $0$ \\
$0$ & $0$ & $0$ & $0$
\end{tabular}
\right),  
\quad && \quad
\lambda_8 = -\frac{1}{\sqrt{3}}\left(
\begin{tabular}{c c c c}
$1$ & $0$ & $0$ & $0$ \\
$0$ & $1$ & $0$ & $0$ \\
$0$ & $0$ & $-2$ & $0$ \\
$0$ & $0$ & $0$ & $0$
\end{tabular}
\right), \nn \\
\lambda_9 = \left(
\begin{tabular}{c c c c}
$0$ & $0$ & $0$ & $1$ \\
$0$ & $0$ & $0$ & $0$ \\
$0$ & $0$ & $0$ & $0$ \\
$1$ & $0$ & $0$ & $0$
\end{tabular}
\right), 
\quad && \quad
\lambda_{10} = \left(
\begin{tabular}{c c c c}
$0$ & $0$ & $0$ & $-i$ \\
$0$ & $0$ & $0$ & $0$ \\
$0$ & $0$ & $0$ & $0$ \\
$i$ & $0$ & $0$ & $0$
\end{tabular}
\right), \nn \\
\lambda_{11} = \left(
\begin{tabular}{c c c c}
$0$ & $0$ & $0$ & $0$ \\
$0$ & $0$ & $0$ & $1$ \\
$0$ & $0$ & $0$ & $0$ \\
$0$ & $1$ & $0$ & $0$
\end{tabular}
\right), 
\quad && \quad
\lambda_{12} = \left(
\begin{tabular}{c c c c}
$0$ & $0$ & $0$ & $0$ \\
$0$ & $0$ & $0$ & $-i$ \\
$0$ & $0$ & $0$ & $0$ \\
$0$ & $i$ & $0$ & $0$
\end{tabular}
\right), \nn \\
\lambda_{13} = \left(
\begin{tabular}{c c c c}
$0$ & $0$ & $0$ & $0$ \\
$0$ & $0$ & $0$ & $0$ \\
$0$ & $0$ & $0$ & $1$ \\
$0$ & $0$ & $1$ & $0$
\end{tabular}
\right), 
\quad && \quad
\lambda_{14} = \left(
\begin{tabular}{c c c c}
$0$ & $0$ & $0$ & $0$ \\
$0$ & $0$ & $0$ & $0$ \\
$0$ & $0$ & $0$ & $i$ \\
$0$ & $0$ & $-i$ & $0$
\end{tabular}
\right), \nn \\
\lambda_{15} = \frac{1}{\sqrt{6}}\left(
\begin{tabular}{c c c c}
$1$ & $0$ & $0$ & $0$ \\
$0$ & $1$ & $0$ & $0$ \\
$0$ & $0$ & $1$ & $0$ \\
$0$ & $0$ & $0$ & $-3$
\end{tabular}
\right), && \nn
\eea 
that they are orthogonal and satisfy $\Tr (\lambda_A)^2=2$, with $A=1,2,\cdots,15$.

\section{$4 \times 4$ $\Phi$ fields}
\label{sec:matrixphi}

\bea
\Phi &=& T_0 \sigma_0 +T_8 \sigma_8 + T_{15} \sigma_{15} = \frac{1}{2} \lambda_0 \sigma_0 + \frac{1}{2} \lambda_8 \sigma_8 + \frac{1}{2} \lambda_{15} \sigma_{15} \nn \\
&=& \frac{1}{2} \sqrt{\frac{1}{2}} \left(
\begin{tabular}{c c c c}
$\sigma_0$ & $0$ & $0$ & $0$ \\
$0$ & $\sigma_0$ & $0$ & $0$ \\
$0$ & $0$ & $\sigma_0$ & $0$ \\
$0$ & $0$ & $0$ & $\sigma_0$
\end{tabular}
\right) 
+\frac{1}{2} \sqrt{\frac{1}{3}} \left(
\begin{tabular}{c c c c}
$\sigma_8$ & $0$ & $0$ & $0$ \\
$0$ & $\sigma_8$ & $0$ & $0$ \\
$0$ & $0$ & $-2\sigma_8$ & $0$ \\
$0$ & $0$ & $0$ & $0$
\end{tabular}
\right) \nn \\
&+& \frac{1}{2} \sqrt{\frac{1}{6}} \left(
\begin{tabular}{c c c c}
$\sigma_{15}$ & $0$ & $0$ & $0$ \\
$0$ & $\sigma_{15}$ & $0$ & $0$ \\
$0$ & $0$ & $-2\sigma_{15}$ & $0$ \\
$0$ & $0$ & $0$ & $-3\sigma_{15}$
\end{tabular}
\right) \nn \\
&=& \frac{1}{2} \left(
\begin{tabular}{c c c c}
$\frac{\sigma_{0}}{\sqrt{2}}+\frac{\sigma_{8}}{\sqrt{3}}+\frac{\sigma_{15}}{\sqrt{6}}$ & $0$ & $0$ & $0$ \\
$0$ & $\frac{\sigma_{0}}{\sqrt{2}}+\frac{\sigma_{8}}{\sqrt{3}}+\frac{\sigma_{15}}{\sqrt{6}}$  & $0$ & $0$ \\
$0$ & $0$ & $\frac{\sigma_{0}}{\sqrt{2}}-\frac{2\sigma_{8}}{\sqrt{3}}+\frac{\sigma_{15}}{\sqrt{6}}$  & $0$ \\
$0$ & $0$ & $0$ & $\frac{\sigma_0}{\sqrt{2}}-\sqrt{\frac{3}{2}}\sigma_{15}$ 
\end{tabular}
\right). \label{eq:phimatrix}
\eea
Similarly, we get {\footnotesize
\bea
\Phi^{\dagger} &=& 
\frac{1}{2} \left(
\begin{tabular}{c c c c}
$\frac{\sigma_{0}}{\sqrt{2}}+\frac{\sigma_{8}}{\sqrt{3}}+\frac{\sigma_{15}}{\sqrt{6}}$ & $0$ & $0$ & $0$ \\
$0$ & $\frac{\sigma_{0}}{\sqrt{2}}+\frac{\sigma_{8}}{\sqrt{3}}+\frac{\sigma_{15}}{\sqrt{6}}$ & $0$ & $0$ \\
$0$ & $0$ & $\sqrt{2}\left(\frac{\sigma_{0}}{2}-\sqrt{\frac{2}{3}}\sigma_{8}+\frac{\sigma_{15}}{2\sqrt{3}}\right)$ & $0$ \\
$0$ & $0$ & $0$ & $\sqrt{2}\left(\frac{\sigma_0}{2}-\frac{\sqrt{3}}{2}\sigma_{15}\right)$ 
\end{tabular}
\right) \label{eq:phidggrmatrix}
\eea
so that
\bea
\Phi \Phi^{\dagger} &=& 
\frac{1}{4} \left(
\begin{tabular}{c c c c}
$\left(\frac{\sigma_{0}}{\sqrt{2}}+\frac{\sigma_{8}}{\sqrt{3}}+\frac{\sigma_{15}}{\sqrt{6}}\right)^2$ & $0$ & $0$ & $0$ \\
$0$ & $\left(\frac{\sigma_{0}}{\sqrt{2}}+\frac{\sigma_{8}}{\sqrt{3}}+\frac{\sigma_{15}}{\sqrt{6}}\right)^2$ & $0$ & $0$ \\
$0$ & $0$ & $2\left(\frac{\sigma_{0}}{2}-\sqrt{\frac{2}{3}}\sigma_{8}+\frac{\sigma_{15}}{2\sqrt{3}}\right)^2$ & $0$ \\
$0$ & $0$ & $0$ & $2\left(\frac{\sigma_0}{2}-\frac{\sqrt{3}}{2}\sigma_{15}\right)^2$ 
\end{tabular}
\right). \hspace*{11mm} \label{eq:phiphidggrmatrix}
\eea
}

\section{Explicit Chiral Symmetry Breaking}
The chiral symmetry in explicitly broken {\footnotesize
\bea
H &=& \sum_{a=0}^{15} h_a T_a = h_0 T_0 + h_8 T_8 + h_{15} T_{15} \nn \\
&=& \frac{1}{2} \left(
\begin{tabular}{c c c c}
$\frac{h_0}{\sqrt{2}}  + h_3 + \frac{h_8}{\sqrt{3}} + \frac{h_{15}}{\sqrt{6}}$ & $h_1 - i h_2$ & $h_4 - i h_5$ & $h_9 - i h_{10}$ \\
$h_1+ih_2$ & $\frac{h_0}{\sqrt{2}}  - h_3 + \frac{h_8}{\sqrt{3}} + \frac{h_{15}}{\sqrt{6}} $ & $h_6-ih_7$ & $h_{11}-i h_{12}$ \\
$h_4+ih_5$ &  $h_6+ih_7$ & $\frac{h_0}{\sqrt{2}}   + 2\frac{h_8}{\sqrt{3}} + \frac{h_{15}}{\sqrt{6}} $ & $h_{13}-i h_{14}$ \\
$h_9-i h_{10}$ &  $h_{11}+ih_{12}$ & $h_{13}+ih_{14}$ & $\frac{h_0}{\sqrt{2}}   + \sqrt{\frac{2}{3}} h_{15} $ 
\end{tabular}
\right). \label{eq:Hmatrix} 
\eea
}

\section{Wavefunction renormalization factors}
\label{sec:wavefrenorfact}

Here, we summarize the wavefunction renormalization factors needed for different meson masses,
\bea
Z_{\pi} = Z_{\eta} = \frac{m_{a_1}}{\sqrt{m_{a_1}^2-g_1^2 \sigma_x^2}},\quad && \quad 
Z_k = \frac{2 m_{k_1}}{\sqrt{4 m_{k_1}^2-g_1^2\left[\sigma_x+\sqrt{2}\sigma_y\right]^2}}, \\
Z_{k_s} = \frac{2m_{K}}{\sqrt{4m_{K}^2-g_1^2 \left[\sigma_x+\sqrt{2}\sigma_y\right]^2}},\quad && \quad
Z_{\eta_s} = \frac{m_{f_{1s}}}{\sqrt{m_{f_{1s}}^2-2g_1^2 \sigma_y^2}}, \\
Z_{\eta_c} = \frac{m_{\chi_{c1}}}{\sqrt{m_{\chi_{c1}}^2 - 2 g1^2 \sigma_c^2}}, \quad && \quad Z_D = \frac{2 m_{D_1}}{\sqrt{4 m_{D_1} - g_1^2 \left(\sigma_c + \sqrt{2}\sigma_c\right)^2}}, \\
Z_{\eta_s} = \frac{m_{f_{1s}}}{\sqrt{m_{f_{1s}}^2 - 2 g_1^2 \sigma_y^2}}, \quad && \quad Z_{D_s} = \frac{\sqrt{2} m_{D_{s_1}}}{\sqrt{2 m_{D_{s_1}}^2 - g_1^2 (\sigma_y^2 + \sigma_c)^2}}, \\ 
Z_{D_0^*} = \frac{2 m_{D^*}}{\sqrt{4 m_{D^*} - g_1^2 (\sigma_x^2 - \sqrt{2}\sigma_c)^2}}, \quad && \quad Z_{D_0^{*0}} = \frac{2 m_{D^{*0}}}{\sqrt{4 m_{D^{*0}}^2 - g_1^2 (\sigma_x - \sqrt{2}\sigma_c)^2}}, \\
Z_{D_{s0}^{*}} = \frac{ \sqrt{2} m_{D^{*}_{s}} }{
\sqrt{2 m_{D^{*}_{s}}^2 - g_1^2 \left(\sigma_y - \sigma_c\right)^2}
}, \quad &&
\eea
where
\bea
g_1^2 &=& \frac{m_{a_1}^2}{f_{\pi}^2 Z_{\pi}^2} \left(1-\frac{1}{Z_{\pi}^2}\right).
\eea

\subsection{Mesonic Potential of SU(4) Linear-Sigma Model}

We start with the mesonic potential of the SU(3) linear-sigma model,
\bea
U_m^{SU(3)}(\Phi) &=& m^2 \Tr \left(\Phi^{\dagger}\Phi\right) + \lambda_1\left[\Tr\left(\Phi^{\dagger}\Phi\right)\right]^2 + \lambda_2\left[\Tr\left(\Phi^{\dagger}\Phi\right)\right]^2 \nn \\
&-& C \left[\det\left(\Phi\right) + \det\left(\Phi^{\dagger}\right)\right] - \Tr\left[H\left(\Phi + \Phi^{\dagger}\right)\right]. \label{eq:SU3Pot}
\eea
It should be noted that Eq. \eqref{eq:SU3Pot} refers to a different anomaly term that the one in the main text. For the current purpose this is just a fine detail.
The mesonic potential of SU(4) has an analogous form as that of SU(3). For a precise estimation of the mass spectrum, a new term must be added, namely $-2\Tr\left[\epsilon \Phi^{\dagger} \Phi\right]$ so that
\bea
U_m^{SU(4)}(\Phi) &=& \frac{m^2}{2} \left(\sigma_x^2+\sigma_y^2+\sigma_{15}^2\right)^2 \nn \\
&+& \lambda_1\left[4\left(\sigma_x+\frac{\sigma_{15}}{\sqrt{6}}\right)^4 + \left(\sqrt{2}\sigma_y+\frac{\sigma_{15}}{\sqrt{6}}\right)^4 + \left(\sqrt{\frac{2}{3}} \sigma_0 - \sqrt{\frac{3}{2}}\sigma_{15}\right)^4 \right. \nn \\
&&\left. + 4\left(\sigma_x+\frac{\sigma_{15}}{\sqrt{6}}\right)^2 \left(\sqrt{2}\sigma_y+\frac{\sigma_{15}}{\sqrt{6}}\right)^2 
+ 4\left(\sigma_x+\frac{\sigma_{15}}{\sqrt{6}}\right)^2   \left(\sqrt{\frac{3}{2}}\sigma_0-\sqrt{\frac{3}{2}} \sigma_{15}\right)^2 \right. \nn \\
&&\left. + 2 \left(\sqrt{2}\sigma_y+\frac{\sigma_{15}}{\sqrt{6}}\right)^2 \left(\sqrt{\frac{2}{3}}\sigma_0-\sqrt{\frac{3}{2}} \sigma_{15}\right)^2
\right]  \\
&+& \lambda_2\left[2\left(\sigma_x+\frac{\sigma_{15}}{\sqrt{6}}\right)^4 + \left(\sqrt{2}\sigma_y+\frac{\sigma_{15}}{\sqrt{6}}\right)^4 + \left(\sqrt{\frac{2}{3}} \sigma_0-\sqrt{\frac{3}{2}}\sigma_{15}\right)^4
\right] \nn \\
&-& \frac{c}{8} \left[\frac{2}{3} \sigma_x^2 \sigma_y \sigma_0 + \frac{\sigma_y \sigma_{15}^2 \sigma_0}{3\sqrt{3}} + \frac{2\sqrt{2}}{3}\sigma_x \sigma_y \sigma_{15} \sigma_0 + \frac{1}{3}\sigma_x^2 \sigma_{15} \sigma_0 + \frac{\sigma_{15}^3 \sigma_0}{8} \right. \nn \\
&& \left. + \frac{\sqrt{2}\sigma_x\sigma_{15}\sigma_0^2}{3\sqrt{3}} - \sqrt{3} \sigma_x^2 \sigma_y \sigma_{15} - \frac{\sigma_{15}^3 \sigma_y}{2\sqrt{3}} - \sqrt{2}\sigma_x\sigma_y\sigma_{15}^2 - \frac{\sigma_x^2\sigma_{15}^2}{2} - \frac{\sigma_{15}^4}{12} - \frac{\sigma_x \sigma_{15}^3}{\sqrt{6}}
\right]. \label{eq:SU4Pot} \nn
\eea

\section{Parameters of the SU(4) Linear-Sigma Model}
\label{sec:Parameters}

We start with the potential of the SU(4) linear-sigma model, Eq. (\ref{eq:SU4Pot}). The global minimum, i.e., vanishing partial derivatives with respect to $\sigma_x$,  $\sigma_y$, and  $\sigma_c$ lead to
\bea
h_x &=& m^2 \sigma_x - \frac{c}{2}  \sigma_x  \sigma_y  \sigma_c + \lambda_1  \sigma_x  \sigma_y^2 + 
\lambda_2  \sigma_x  \sigma_c^2 + \frac{1}{2}\left(2\lambda_1+\lambda_2\right)  \sigma_x^2, \\
h_y &=& m^2 \sigma_y - \frac{c}{2}  \sigma_x^2  \sigma_c + \lambda_1  \sigma_x^2  \sigma_y + 
\lambda_2  \sigma_y  \sigma_c^2 + \left(\lambda_1+\lambda_2\right)  \sigma_y^2, \\
h_c &=& m^2 \sigma_c - \frac{c}{2}  \sigma_x^2  \sigma_y  + \lambda_1  \sigma_x^2  \sigma_c + 
\lambda_2  \sigma_y^2  \sigma_c + \left(\lambda_1+\lambda_2\right)  \sigma_c^2,
\eea
where $\lambda_1$ and $\lambda_2$ are respectively defined as
\bea
\lambda_1 &=& \frac{m_{\sigma}^2 - m_{\pi}^2 - m_{a_0}^2 + m^2_{\eta}}{3f_{\pi}^2}, \\
\lambda_2 &=& \frac{3\left(2f_{K}-f_{\pi}\right)m_K^2 - 3\left(2f_{K}-f_{\pi}\right)m_{\pi}^2 - 2\left(f_{K}-f_{\pi}\right)\left(m_{\eta^{\prime}}^2 + m_{\eta}^2\right)}{\left(f_{K}-f_{\pi}\right)\left(3f_{\pi}^2+8f_K\left(f_K-f_{\pi}\right)\right)},
\eea 
where $f_{K}$ and $f_{\pi}$ are the decay constants of $K$ and $\pi$ mesons which can be taken from the recent compilation of the particle data group \cite{10.1093/ptep/ptac097}.

The U(1)$_{\mathtt{A}}$ anomaly breaking term  $C$ is fixed by $\lambda_2$ and the difference of pion and Kaon masses,
\bea
C &=& \frac{m_K^2-m_{\pi}^2}{f_K-f_{\pi}} - \lambda_2\left(2f_K-f_{\pi}\right).
\eea 

The external field $\Delta$ could be expressed from the Lagrangian term $\Tr[\Delta(L^{\mu\nu}+L^{\mu\nu})]$ 
\bea
\Delta &=& \left(
\begin{tabular}{c c c c}
$\delta_u$ & $0$ & $0$ & $0$ \\
$0$ & $\delta_d$ & $0$ & $0$ \\
$0$ & $0$ & $\delta_s$ & $0$ \\
$0$ & $0$ & $0$ & $\delta_c$
\end{tabular}
\right),
\eea
from which we deduce that
\bea
 \left(
\begin{tabular}{c}
$\delta_u$ \\
$\delta_d$  \\
$\delta_s$  \\
$\delta_c$
\end{tabular}
\right) &=& \left(
\begin{tabular}{c}
$m_u^2$ \\
$m_d^2$  \\
$m_s^2$  \\
$m_c^2$
\end{tabular}
\right).
\eea
In the isospin-symmetric limit, it turns possible to set $\delta_u=\delta_d=0$. In this case, for $\delta_x$, $\delta_y$, and $\delta_c$, we might use the mass equations of vector mesons, for example, $m^2_{\omega_N}$, $m^2_{\omega_S}$, and $m^2_{\chi_{c1}}$. Then, we get
\bea
\delta_x &=& \frac{1}{2} \left[m^2_{\omega_N}-m_1^2+m_0^2-\frac{\sigma_x^2}{2}\left(h_1+h_2+h_3\right) -\frac{h_1}{2} \left(\sigma_y^2+\sigma_c^2\right) \right], \\ 
\delta_y &=& \frac{1}{2} \left[m^2_{\omega_S}-m_1^2+m_0^2-\frac{\sigma_y^2}{2}\left(\frac{h_1}{2}+h_2+h_3\right) - \frac{h_1}{2} \left(\sigma_x^2+\sigma_c^2\right) \right], \\ 
\delta_c &=& \frac{1}{2} \left[m^2_{\chi_{c1}}-m_1^2+m_0^2 - 2 g_1^2 \sigma_c^2 - \sigma_c^2\left(\frac{h_1}{2}+h_2-h_3\right) -\frac{h_1}{2} \left(\sigma_y^2+\sigma_y^2\right) \right]. \\ 
\eea

For the mass parameters,
\bea
m^2 &=& m_{\pi}^2 - \frac{f_{\pi}^2}{2} \lambda_2 + \frac{c}{2}\left(2f_K-f_{\phi}\right)- \lambda_1\left(\frac{1}{2}\left(2f_K-f_{\pi}\right)^2\right), \\
m^2_0 &=& \frac{1}{2} \left[m_{a_0}^2+ m_{\sigma_s}^2 - \lambda_2\left(\frac{3}{2} \sigma_x^2 + 3 \sigma_y^2\right)\right], \\
m^2_1 &=& m_{\omega_s}^2 - \left(\frac{h_1}{2}+h_2+h_3\right) \sigma_s^2 - \frac{h_1}{2} \sigma_x^2 - 2 \delta_s.
\eea

Finally, for the parameters $h_1$, $h_2$. and $h_3$, we recall the potential of the SU(3) linear-sigma model, Eq. (\ref{eq:SU3Pot}), where $\sigma_u$, $\sigma_d$, and $\sigma_s$ can be related to $\sigma_0$, $\sigma_3$, and $\sigma_8$,
\bea
\sigma_u &=& \sqrt{2} \sigma_0 + \sigma_3 + \sigma_8, \\
\sigma_d &=& \sqrt{2} \sigma_0 - \sigma_3 + \sigma_8, \\
\sigma_s &=& \sigma_0 - \sqrt{2} \sigma_8.
\eea
Then, $h_0$, $h_3$, and $h_8$ become
\bea
h_0  &=& \frac{1}{\sqrt{6}} \left[f_{\pi} m_{\pi}^2 + 2 f_K m_k^2\right], \\
h_3  &=& \left[m^2 + \frac{c}{\sqrt{6}} \sigma_0 -  \frac{c}{\sqrt{6}} \sigma_8 + \lambda_1 \left(\sigma_0^2+\sigma_3^2+\sigma_8^2\right) \right. \nn \\
&+&\left. \lambda_2 \left(\sigma_0^2 + \frac{\sigma_3^2}{2} +  \frac{\sigma_8^2}{2} + \sqrt{2} \sigma_0 \sigma_8\right)
\right] \sigma_3, \\
h_8  &=& \frac{2}{3} \left[f_{\pi} m_{\pi}^2 - 2 f_K m_k^2\right].  
\eea



\section*{Acknowledgment}

AT acknowledges the generous support by the Egyptian Center for Theoretical Physics (ECTP) and Future University in Egypt (FUE)!

\section*{Conflicts of Interest}

The authors declare that there are no conflicts of interest regarding the publication of this published article!


\section*{Dataset Availability}

All data generated or analyzed during this study are included in this published article. The data used to support the findings of this study are included within the published article and properly cited! All of the material is owned by the authors.

\section*{Author contributions}

The first author was responsible for deriving the analytical expressions. The second author contributed to the writing and proofreading of the manuscript. The responsibility for proposing the conception of the present study lies with the third, who also undertook the tasks of designing and managing the research, interpreting the results, deriving the expressions, drawing the figures, and preparing the manuscript. The final version of the manuscript was unanimously approved by all authors.

\section*{Funding}

The authors declare that this research received no specific grants from any funding agency in the public, commercial, or not-for-profit sectors.


\section*{Competing interests}

The authors confirm that there are no relevant financial or non-financial competing interests to report.

\reftitle{References}


\bibliography{Azar-CharmedMesonStates1}

\begin{thebibliography}{999}

\bibitem[Gell-Mann and Levy(1960)]{Gell-Mann:1960mvl}
Gell-Mann, M.; Levy, M.
\newblock {The axial vector current in beta decay}.
\newblock {\em Nuovo Cim.} {\bf 1960}, {\em 16},~705.
\newblock {\url{https://doi.org/10.1007/BF02859738}}.

\bibitem[Lenaghan and Rischke(2000)]{Lenaghan:1999si}
Lenaghan, J.T.; Rischke, D.H.
\newblock {The O(N) model at finite temperature: Renormalization of the gap
  equations in Hartree and large N approximation}.
\newblock {\em J. Phys. G} {\bf 2000}, {\em 26},~431--450,
  \href{http://arxiv.org/abs/nucl-th/9901049}{{\normalfont [nucl-th/9901049]}}.
\newblock {\url{https://doi.org/10.1088/0954-3899/26/4/309}}.

\bibitem[Petropoulos(1999)]{Petropoulos:1998gt}
Petropoulos, N.
\newblock {Linear sigma model and chiral symmetry at finite temperature}.
\newblock {\em J. Phys. G} {\bf 1999}, {\em 25},~2225--2241,
  \href{http://arxiv.org/abs/hep-ph/9807331}{{\normalfont [hep-ph/9807331]}}.
\newblock {\url{https://doi.org/10.1088/0954-3899/25/11/305}}.

\bibitem[Hu(1974)]{PhysRevD.9.1825}
Hu, B.
\newblock Chiral SU(4) and scale invariance.
\newblock {\em Phys. Rev. D} {\bf 1974}, {\em 9},~1825--1834.
\newblock {\url{https://doi.org/10.1103/PhysRevD.9.1825}}.

\bibitem[Schechter and Singer(1975)]{PhysRevD.12.2781}
Schechter, J.; Singer, M.
\newblock SU(4) sigma model.
\newblock {\em Phys. Rev. D} {\bf 1975}, {\em 12},~2781--2790.
\newblock {\url{https://doi.org/10.1103/PhysRevD.12.2781}}.

\bibitem[Geddes(1980)]{PhysRevD.21.278}
Geddes, H.B.
\newblock Spin-zero mass spectrum in the one-loop approximation in a linear
  SU(4) sigma model.
\newblock {\em Phys. Rev. D} {\bf 1980}, {\em 21},~278--289.
\newblock {\url{https://doi.org/10.1103/PhysRevD.21.278}}.

\bibitem[Andrianov et~al.(2020)Andrianov, Andrianov, and
  Espriu]{particles3010002}
Andrianov, A.; Andrianov, V.; Espriu, D.
\newblock Chiral Perturbation Theory vs. Linear Sigma Model in a Chiral
  Imbalance Medium.
\newblock {\em Particles} {\bf 2020}, {\em 3},~15--22.

\bibitem[Tawfik(2005)]{Tawfik:2004sw}
Tawfik, A.
\newblock {QCD phase diagram: A Comparison of lattice and hadron resonance gas
  model calculations}.
\newblock {\em Phys. Rev. D} {\bf 2005}, {\em 71},~054502,
  \href{http://arxiv.org/abs/hep-ph/0412336}{{\normalfont [hep-ph/0412336]}}.
\newblock {\url{https://doi.org/10.1103/PhysRevD.71.054502}}.

\bibitem[Geddes(1980)]{Geddes:1979nd}
Geddes, H.B.
\newblock {The Spin 0 Mass Spectrum in the One Loop Approximation in a Linear
  SU(4) Sigma Model}.
\newblock {\em Phys. Rev. D} {\bf 1980}, {\em 21},~278.
\newblock {\url{https://doi.org/10.1103/PhysRevD.21.278}}.

\bibitem[Eshraim et~al.(2015)Eshraim, Giacosa, and Rischke]{Eshraim2015}
Eshraim, W.I.; Giacosa, F.; Rischke, D.H.
\newblock Phenomenology of charmed mesons in the extended Linear Sigma Model.
\newblock {\em Eur. Phys. J. A} {\bf 2015}, {\em 51},~112,
  \href{http://arxiv.org/abs/1405.5861 [hep-ph]}{{\normalfont [1405.5861
  [hep-ph]]}}.

\bibitem[Diab et~al.(2016)Diab, Ahmadov, Tawfik, and Dahab]{Diab:2016iig}
Diab, A.M.; Ahmadov, A.I.; Tawfik, A.N.; Dahab, E.A.E.
\newblock {SU($4$) Polyakov linear-sigma model at finite temperature and
  density}.
\newblock {\em PoS} {\bf 2016}, {\em ICHEP2016},~634,
  \href{http://arxiv.org/abs/1611.02693}{{\normalfont
  [arXiv:nucl-th/1611.02693]}}.
\newblock {\url{https://doi.org/10.22323/1.282.0634}}.

\bibitem[Tawfik et~al.(2014)Tawfik, Magdy, and Diab]{Tawfik:2014uka}
Tawfik, A.; Magdy, N.; Diab, A.
\newblock {Polyakov linear SU(3) $\sigma$ model: Features of higher-order
  moments in a dense and thermal hadronic medium}.
\newblock {\em Phys. Rev. C} {\bf 2014}, {\em 89},~055210,
  \href{http://arxiv.org/abs/1405.0577}{{\normalfont
  [arXiv:hep-ph/1405.0577]}}.
\newblock {\url{https://doi.org/10.1103/PhysRevC.89.055210}}.

\bibitem[Tawfik et~al.(2016{\natexlab{a}})Tawfik, Diab, and
  Hussein]{Tawfik:2016edq}
Tawfik, A.N.; Diab, A.M.; Hussein, M.T.
\newblock {SU(3) Polyakov linear-sigma model: Conductivity and viscous
  properties of QCD matter in thermal medium}.
\newblock {\em Int. J. Mod. Phys. A} {\bf 2016}, {\em 31},~1650175,
  \href{http://arxiv.org/abs/1610.06041}{{\normalfont
  [arXiv:nucl-th/1610.06041]}}.
\newblock {\url{https://doi.org/10.1142/S0217751X1650175X}}.

\bibitem[Tawfik et~al.(2016{\natexlab{b}})Tawfik, Diab, Ezzelarab, and
  Shalaby]{Tawfik:2016lih}
Tawfik, A.N.; Diab, A.M.; Ezzelarab, N.; Shalaby, A.G.
\newblock {QCD thermodynamics and magnetization in nonzero magnetic field}.
\newblock {\em Adv. High Energy Phys.} {\bf 2016}, {\em 2016},~1381479,
  \href{http://arxiv.org/abs/1604.00043}{{\normalfont
  [arXiv:hep-ph/1604.00043]}}.
\newblock {\url{https://doi.org/10.1155/2016/1381479}}.

\bibitem[Tawfik et~al.(2018)Tawfik, Diab, and Hussein]{Tawfik:2016gye}
Tawfik, A.N.; Diab, A.M.; Hussein, M.T.
\newblock {Quark\textendash{}hadron phase structure, thermodynamics, and
  magnetization of QCD matter}.
\newblock {\em J. Phys. G} {\bf 2018}, {\em 45},~055008,
  \href{http://arxiv.org/abs/1604.08174}{{\normalfont
  [arXiv:hep-lat/1604.08174]}}.
\newblock {\url{https://doi.org/10.1088/1361-6471/aaba9e}}.

\bibitem[Tawfik et~al.(2019)Tawfik, Diab, and Hussein]{Tawfik:2019rdd}
Tawfik, A.N.; Diab, A.M.; Hussein, M.T.
\newblock {Chiral phase structure of the sixteen meson states in the SU(3)
  Polyakov linear-sigma model for finite temperature and chemical potential in
  a strong magnetic field}.
\newblock {\em Chin. Phys. C} {\bf 2019}, {\em 43},~034103,
  \href{http://arxiv.org/abs/1901.03293}{{\normalfont
  [arXiv:hep-ph/1901.03293]}}.
\newblock {\url{https://doi.org/10.1088/1674-1137/43/3/034103}}.

\bibitem[Tawfik and Diab(2021)]{Tawfik:2021eeb}
Tawfik, A.N.; Diab, A.M.
\newblock {Chiral magnetic properties of QCD phase-diagram}.
\newblock {\em Eur. Phys. J. A} {\bf 2021}, {\em 57},~200,
  \href{http://arxiv.org/abs/2106.04576}{{\normalfont
  [arXiv:hep-ph/2106.04576]}}.
\newblock {\url{https://doi.org/10.1140/epja/s10050-021-00501-z}}.

\bibitem[Tawfik(2023)]{Tawfik:2023egf}
Tawfik, A.N.
\newblock {QCD Phase Structure and In-Medium Modifications of Meson Masses in
  Polyakov Linear-Sigma Model with Finite Isospin Asymmetry}.
\newblock {\em Universe} {\bf 2023}, {\em 9},~276.
\newblock {\url{https://doi.org/10.3390/universe9060276}}.

\bibitem[Tawfik and Magdy(2015)]{Tawfik:2015tga}
Tawfik, A.N.; Magdy, N.
\newblock {SU(3) Polyakov linear-\ensuremath{\sigma} model in magnetic fields:
  Thermodynamics, higher-order moments, chiral phase structure, and meson
  masses}.
\newblock {\em Phys. Rev. C} {\bf 2015}, {\em 91},~015206,
  \href{http://arxiv.org/abs/1501.01124}{{\normalfont
  [arXiv:hep-ph/1501.01124]}}.
\newblock {\url{https://doi.org/10.1103/PhysRevC.91.015206}}.

\bibitem[Tawfik and Diab(2015)]{Tawfik:2014gga}
Tawfik, A.N.; Diab, A.M.
\newblock {Polyakov SU(3) extended linear- \ensuremath{\sigma} model: Sixteen
  mesonic states in chiral phase structure}.
\newblock {\em Phys. Rev. C} {\bf 2015}, {\em 91},~015204,
  \href{http://arxiv.org/abs/1412.2395}{{\normalfont
  [arXiv:hep-ph/1412.2395]}}.
\newblock {\url{https://doi.org/10.1103/PhysRevC.91.015204}}.

\bibitem[Tawfik et~al.(2019)Tawfik, Diab, Ghoneim, and Anwer]{Tawfik:2019tkp}
Tawfik, A.N.; Diab, A.M.; Ghoneim, M.T.; Anwer, H.
\newblock {SU(3) Polyakov Linear-Sigma Model With Finite Isospin Asymmetry: QCD
  Phase Diagram}.
\newblock {\em Int. J. Mod. Phys. A} {\bf 2019}, {\em 34},~1950199,
  \href{http://arxiv.org/abs/1904.09890}{{\normalfont
  [arXiv:hep-ph/1904.09890]}}.
\newblock {\url{https://doi.org/10.1142/S0217751X19501999}}.

\bibitem[Tawfik et~al.(2020)Tawfik, Greiner, Diab, Ghoneim, and
  Anwer]{Tawfik:2019kaz}
Tawfik, A.N.; Greiner, C.; Diab, A.M.; Ghoneim, M.T.; Anwer, H.
\newblock {Polyakov linear-$\sigma$ model in mean-field approximation and
  optimized perturbation theory}.
\newblock {\em Phys. Rev. C} {\bf 2020}, {\em 101},~035210,
  \href{http://arxiv.org/abs/1908.05939}{{\normalfont
  [arXiv:hep-ph/1908.05939]}}.
\newblock {\url{https://doi.org/10.1103/PhysRevC.101.035210}}.

\bibitem[Giacosa et~al.(2012)Giacosa, Parganlija, Kovacs, and
  Wolf]{Giacosa:2012cq}
Giacosa, F.; Parganlija, D.; Kovacs, P.; Wolf, G.
\newblock {Phenomenology of light mesons within a chiral approach}.
\newblock {\em EPJ Web Conf.} {\bf 2012}, {\em 37},~08006,
  \href{http://arxiv.org/abs/1208.6202}{{\normalfont
  [arXiv:hep-ph/1208.6202]}}.
\newblock {\url{https://doi.org/10.1051/epjconf/20123708006}}.

\bibitem[Tawfik(2023)]{Tawfik:2023all}
Tawfik, A.N.
\newblock {Isospin Symmetry Breaking in Non-Perturbative QCD}.
\newblock {\em Phys. Sci. Forum} {\bf 2023}, {\em 7},~22.
\newblock {\url{https://doi.org/10.3390/ECU2023-14047}}.

\bibitem[Tawfik and Magdy(2015)]{Tawfik:2015uda}
Tawfik, A.N.; Magdy, N.
\newblock {On SU(3) effective models and chiral phase-transition}.
\newblock {\em Adv. High Energy Phys.} {\bf 2015}, {\em 2015},~563428,
  \href{http://arxiv.org/abs/1509.07114}{{\normalfont
  [arXiv:hep-ph/1509.07114]}}.
\newblock {\url{https://doi.org/10.1155/2015/563428}}.

\bibitem[Eshraim et~al.(2020)Eshraim, Fischer, Giacosa, and
  Parganlija]{Eshraim2020}
Eshraim, W.I.; Fischer, C.S.; Giacosa, F.; Parganlija, D.
\newblock Hybrid phenomenology in a chiral approach.
\newblock {\em Eur. Phys. J. Plus} {\bf 2020}, {\em 135},~945,
  \href{http://arxiv.org/abs/2001.06106 [hep-ph]}{{\normalfont [2001.06106
  [hep-ph]]}}.

\bibitem[Klempt and Zaitsev(2007)]{Klempt:2007cp}
Klempt, E.; Zaitsev, A.
\newblock {Glueballs, Hybrids, Multiquarks. Experimental facts versus QCD
  inspired concepts}.
\newblock {\em Phys. Rept.} {\bf 2007}, {\em 454},~1--202,
  \href{http://arxiv.org/abs/0708.4016}{{\normalfont
  [arXiv:hep-ph/0708.4016]}}.
\newblock {\url{https://doi.org/10.1016/j.physrep.2007.07.006}}.

\bibitem[Amsler and Törnqvist(2004)]{AMSLER200461}
Amsler, C.; Törnqvist, N.A.
\newblock Mesons beyond the naive quark model.
\newblock {\em Physics Reports} {\bf 2004}, {\em 389},~61--117.
\newblock
  {\url{https://doi.org/https://doi.org/10.1016/j.physrep.2003.09.003}}.

\bibitem[Polyakov(1978)]{POLYAKOV1978477}
Polyakov, A.
\newblock Thermal properties of gauge fields and quark liberation.
\newblock {\em Physics Letters B} {\bf 1978}, {\em 72},~477--480.
\newblock {\url{https://doi.org/https://doi.org/10.1016/0370-2693(78)90737-2}}.

\bibitem[Susskind(1979)]{PhysRevD.20.2610}
Susskind, L.
\newblock Lattice models of quark confinement at high temperature.
\newblock {\em Phys. Rev. D} {\bf 1979}, {\em 20},~2610--2618.
\newblock {\url{https://doi.org/10.1103/PhysRevD.20.2610}}.

\bibitem[Svetitsky and Yaffe(1982)]{SVETITSKY1982423}
Svetitsky, B.; Yaffe, L.G.
\newblock Critical behavior at finite-temperature confinement transitions.
\newblock {\em Nuclear Physics B} {\bf 1982}, {\em 210},~423--447.
\newblock {\url{https://doi.org/https://doi.org/10.1016/0550-3213(82)90172-9}}.

\bibitem[Vafa and Witten(1984)]{VAFA1984173}
Vafa, C.; Witten, E.
\newblock Restrictions on symmetry breaking in vector-like gauge theories.
\newblock {\em Nuclear Physics B} {\bf 1984}, {\em 234},~173--188.
\newblock {\url{https://doi.org/https://doi.org/10.1016/0550-3213(84)90230-X}}.

\bibitem[Workman et~al.(2022)]{ParticleDataGroup:2022pth}
Workman, R.L.;  et~al.
\newblock {Review of Particle Physics}.
\newblock {\em PTEP} {\bf 2022}, {\em 2022},~083C01.
\newblock {\url{https://doi.org/10.1093/ptep/ptac097}}.

\bibitem[Barbieri et~al.(1976)Barbieri, Kögerler, Kunszt, and
  Gatto]{BARBIERI1976125}
Barbieri, R.; Kögerler, R.; Kunszt, Z.; Gatto, R.
\newblock Meson masses and widths in a gauge theory with linear binding
  potential.
\newblock {\em Nuclear Physics B} {\bf 1976}, {\em 105},~125--138.
\newblock {\url{https://doi.org/https://doi.org/10.1016/0550-3213(76)90064-X}}.

\bibitem[Parganlija et~al.(2011)Parganlija, Giacosa, Kovacs, and
  Wolf]{Parganlija:2010ac}
Parganlija, D.; Giacosa, F.; Kovacs, P.; Wolf, G.
\newblock {Masses of Axial-Vector Resonances in a Linear Sigma Model with
  Nf=3}.
\newblock {\em AIP Conf. Proc.} {\bf 2011}, {\em 1343},~328--330,
  \href{http://arxiv.org/abs/1011.6104}{{\normalfont
  [arXiv:hep-ph/1011.6104]}}.
\newblock {\url{https://doi.org/10.1063/1.3575019}}.

\bibitem[Fariborz et~al.(2009)Fariborz, Park, Schechter, and
  Shahid]{PhysRevD.80.113001}
Fariborz, A.H.; Park, N.W.; Schechter, J.; Shahid, M.N.
\newblock Gauged linear sigma model and pion-pion scattering.
\newblock {\em Phys. Rev. D} {\bf 2009}, {\em 80},~113001.
\newblock {\url{https://doi.org/10.1103/PhysRevD.80.113001}}.

\bibitem[Parganlija et~al.(2010)Parganlija, Giacosa, and
  Rischke]{Parganlija:2010fz}
Parganlija, D.; Giacosa, F.; Rischke, D.H.
\newblock {Vacuum Properties of Mesons in a Linear Sigma Model with Vector
  Mesons and Global Chiral Invariance}.
\newblock {\em Phys. Rev. D} {\bf 2010}, {\em 82},~054024,
  \href{http://arxiv.org/abs/1003.4934}{{\normalfont
  [arXiv:hep-ph/1003.4934]}}.
\newblock {\url{https://doi.org/10.1103/PhysRevD.82.054024}}.

\bibitem[Napsuciale(1998)]{Napsuciale:1998ip}
Napsuciale, M.
\newblock {Scalar meson masses and mixing angle in a U(3) x U(3) linear sigma
  model} {\bf 1998}.
\newblock  \href{http://arxiv.org/abs/hep-ph/9803396}{{\normalfont
  [hep-ph/9803396]}}.

\bibitem[Gasser and Leutwyler(1984)]{GASSER1984142}
Gasser, J.; Leutwyler, H.
\newblock Chiral perturbation theory to one loop.
\newblock {\em Annals of Physics} {\bf 1984}, {\em 158},~142--210.
\newblock {\url{https://doi.org/https://doi.org/10.1016/0003-4916(84)90242-2}}.

\bibitem[Bando et~al.(1988)Bando, Kugo, and Yamawaki]{Bando:1987br}
Bando, M.; Kugo, T.; Yamawaki, K.
\newblock {Nonlinear Realization and Hidden Local Symmetries}.
\newblock {\em Phys. Rept.} {\bf 1988}, {\em 164},~217--314.
\newblock {\url{https://doi.org/10.1016/0370-1573(88)90019-1}}.

\bibitem[Rosenzweig et~al.(1981)Rosenzweig, Salomone, and
  Schechter]{Rosenzweig:1981cu}
Rosenzweig, C.; Salomone, A.; Schechter, J.
\newblock {A Pseudoscalar Glueball, the Axial Anomaly and the Mixing Problem
  for Pseudoscalar Mesons}.
\newblock {\em Phys. Rev. D} {\bf 1981}, {\em 24},~2545--2548.
\newblock {\url{https://doi.org/10.1103/PhysRevD.24.2545}}.

\bibitem[Parganlija et~al.(2013)Parganlija, Kovacs, Wolf, Giacosa, and
  Rischke]{Parganlija:2012fy}
Parganlija, D.; Kovacs, P.; Wolf, G.; Giacosa, F.; Rischke, D.H.
\newblock {Meson vacuum phenomenology in a three-flavor linear sigma model with
  (axial-)vector mesons}.
\newblock {\em Phys. Rev. D} {\bf 2013}, {\em 87},~014011,
  \href{http://arxiv.org/abs/1208.0585}{{\normalfont
  [arXiv:hep-ph/1208.0585]}}.
\newblock {\url{https://doi.org/10.1103/PhysRevD.87.014011}}.

\bibitem[Mott(1975)]{MOTT1975260}
Mott, R.
\newblock SU (4) × SU (4) symmetry breaking and its implications for SU (3) ×
  SU (3) breaking.
\newblock {\em Nuclear Physics B} {\bf 1975}, {\em 84},~260--268.
\newblock {\url{https://doi.org/https://doi.org/10.1016/0550-3213(75)90550-7}}.

\bibitem[Myers and Ogilvie(2008)]{PhysRevD.77.125030}
Myers, J.C.; Ogilvie, M.C.
\newblock New phases of $SU(3)$ and $SU(4)$ at finite temperature.
\newblock {\em Phys. Rev. D} {\bf 2008}, {\em 77},~125030.
\newblock {\url{https://doi.org/10.1103/PhysRevD.77.125030}}.

\bibitem[Tawfik(2014)]{Tawfik:2014eba}
Tawfik, A.N.
\newblock {Equilibrium statistical-thermal models in high-energy physics}.
\newblock {\em Int. J. Mod. Phys. A} {\bf 2014}, {\em 29},~1430021,
  \href{http://arxiv.org/abs/1410.0372}{{\normalfont
  [arXiv:hep-ph/1410.0372]}}.
\newblock {\url{https://doi.org/10.1142/S0217751X1430021X}}.

\bibitem[Feng(2008)]{Feng:2008zz}
Feng, X.C.
\newblock {On the mass spectrum of 2(1)S0 meson state}.
\newblock {\em Acta Phys. Polon. B} {\bf 2008}, {\em 39},~2931--2938.

\bibitem[Freund(1968)]{Freund1968}
Freund, P.
\newblock Radially excited mesons.
\newblock {\em Nuovo Cimento A} {\bf 1968}, {\em 58},~519--523.
\newblock {\url{https://doi.org/https://doi.org/10.1007/BF02819151}}.

\bibitem[Chen-Tsai and Lee(1974)]{PhysRevD.10.2960}
Chen-Tsai, C.T.; Lee, T.Y.
\newblock Radially excited states of mesons.
\newblock {\em Phys. Rev. D} {\bf 1974}, {\em 10},~2960--2962.
\newblock {\url{https://doi.org/10.1103/PhysRevD.10.2960}}.

\bibitem[Sbaih et~al.(2013)Sbaih, Srour, and Fayad]{Sbaih2013}
Sbaih, M.; Srour, M.; Fayad, H.
\newblock Lie Algebra and Representation of SU(4).
\newblock {\em Elecr. J. Theor. Phys.} {\bf 2013}, {\em 10},~9--26.

\bibitem[Arfken and Weber(1985)]{Arfken1985}
Arfken, G.; Weber, H.J.
\newblock {\em Mathematical Methods for Physicists}; Elsevier Academic Press,
  California, USA,  1985.

\bibitem[Group et~al.(2022)Group, Workman, Burkert, Crede, Klempt, Thoma,
  Tiator, Agashe, Aielli, Allanach, Amsler, Antonelli, Aschenauer, Asner, Baer,
  Banerjee, Barnett, Baudis, Bauer, Beatty, Belousov, Beringer, Bettini,
  Biebel, Black, Blucher, Bonventre, Bryzgalov, Buchmuller, Bychkov, Cahn,
  Carena, Ceccucci, Cerri, Chivukula, Cowan, Cranmer, Cremonesi, D'Ambrosio,
  Damour, de~Florian, de~GouvÃªa, DeGrand, de~Jong, Demers, Dobrescu,
  D'Onofrio, Doser, Dreiner, Eerola, Egede, Eidelman, El-Khadra, Ellis, Eno,
  Erler, Ezhela, Fetscher, Fields, Freitas, Gallagher, Gershtein, Gherghetta,
  Gonzalez-Garcia, Goodman, Grab, Gritsan, Grojean, Groom, GrÃ¼newald, Gurtu,
  Gutsche, Haber, Hamel, Hanhart, Hashimoto, Hayato, Hebecker, Heinemeyer,
  HernÃ¡ndez-Rey, Hikasa, Hisano, HÃ¶cker, Holder, Hsu, Huston, Hyodo,
  Ianni, Kado, Karliner, Katz, Kenzie, Khoze, Klein, Krauss, Kreps, KriÅ¾an,
  Krusche, Kwon, Lahav, Laiho, Lellouch, Lesgourgues, Liddle, Ligeti, Lin,
  Lippmann, Liss, Littenberg, LourenÃ§o, Lugovsky, Lugovsky, Lusiani, Makida,
  Maltoni, Mannel, Manohar, Marciano, Masoni, Matthews, MeiÃŸner,
  Melzer-Pellmann, Mikhasenko, Miller, Milstead, Mitchell, MÃ¶nig, Molaro,
  Moortgat, Moskovic, Nakamura, Narain, Nason, Navas, Nelles, Neubert, Nevski,
  Nir, Olive, Patrignani, Peacock, Petrov, Pianori, Pich, Piepke, Pietropaolo,
  Pomarol, Pordes, Profumo, Quadt, Rabbertz, Rademacker, Raffelt,
  Ramsey-Musolf, Ratcliff, Richardson, Ringwald, Robinson, Roesler, Rolli,
  Romaniouk, Rosenberg, Rosner, Rybka, Ryskin, Ryutin, Sakai, Sarkar, Sauli,
  Schneider, SchÃ¶nert, Scholberg, Schwartz, Schwiening, Scott, Sefkow,
  Seljak, Sharma, Sharpe, Shiltsev, Signorelli, Silari, Simon, SjÃ¶strand,
  Skands, Skwarnicki, Smoot, Soffer, Sozzi, Spanier, Spiering, Stahl, Stone,
  Sumino, Syphers, Takahashi, Tanabashi, Tanaka, TaÅ¡evskÃ½, Terao,
  Terashi, Terning, Thorne, Titov, Tkachenko, Tovey, Trabelsi, Urquijo,
  Valencia, Van~de Water, Varelas, Venanzoni, Verde, Vivarelli, Vogel,
  Vogelsang, Vorobyev, Wakely, Walkowiak, Walter, Wands, Weinberg, Weinberg,
  Wermes, White, Wiencke, Willocq, Wohl, Woody, Yao, Yokoyama, Yoshida,
  Zanderighi, Zeller, Zenin, Zhu, Zhu, Zimmermann, and
  Zyla]{10.1093/ptep/ptac097}
Group, P.D.; Workman, R.L.; Burkert, V.D.; Crede, V.; Klempt, E.; Thoma, U.;
  Tiator, L.; Agashe, K.; Aielli, G.; Allanach, B.C.;  et~al.
\newblock {Review of Particle Physics}.
\newblock {\em Progress of Theoretical and Experimental Physics} {\bf 2022},
  {\em 2022},~083C01.
\newblock {\url{https://doi.org/10.1093/ptep/ptac097}}.

\end{thebibliography}


%



\end{document}